\documentclass[aps,twocolumn,showpacs,showkeys]{revtex4}
\usepackage{graphicx,subfigure,mathrsfs,amsfonts}

\newcommand{\bastar}{\begin{eqnarray*}}
\newcommand{\eastar}{\end{eqnarray*}}
\newskip\humongous \humongous=0pt plus 1000pt minus 1000pt

\newif\ifdtup

\relax
\newcommand{\bea}{\begin{eqnarray}}
\newcommand{\eea}{\end{eqnarray}}
\newcommand{\nn}{\nonumber}
\newcommand{\pro}{\partial}
\newcommand{\oneg}{\displaystyle\frac{1}{g}}
\newcommand{\dfrac}{\displaystyle\frac}
\newcommand{\mn}{{\mu\nu}}
\newcommand{\A}{{\vec A}}
\newcommand{\bA}{{\bar A}}

\newcommand{\hA}{{\hat A}}
\newcommand{\B}{{\vec B}}

\newcommand{\hB}{{\hat B}}
\newcommand{\tB}{{\widetilde{B}}}
\newcommand{\C}{{\vec C}}

\newcommand{\tC}{{\widetilde{C}}}
\newcommand{\hD}{{\hat D}}
\newcommand{\bD}{{\bar D}}
\newcommand{\cD}{{\cal D}}

\newcommand{\E}{{\vec E}}
\newcommand{\e}{{\vec e}}
\newcommand{\he}{{\hat e}}

\newcommand{\F}{{\vec F}}
\newcommand{\bF}{{\bar F}}

\newcommand{\hF}{{\hat F}}
\newcommand{\vg}{{\bf g}}
\newcommand{\hvg}{\hat {{\bf g}}}
\newcommand{\vG}{{\bf G}}

\newcommand{\tG}{{\widetilde{G}}}
\newcommand{\Gm}{{\Gamma}}

\newcommand{\vGm}{{\bf \Gamma}}
\newcommand{\hvGm}{\hat{{\bf \Gamma}}}
\newcommand{\tGm}{{\widetilde{\Gamma}}}
\newcommand{\tH}{{\widetilde{H}}}
\newcommand{\vI}{{\bf I}}
\newcommand{\vj}{{\bf {j}}}
\newcommand{\tj}{\tilde{j}}
\newcommand{\vtj}{\tilde{{\bf j}}}
\newcommand{\tJ}{{\widetilde{J}}}
\newcommand{\vk}{{\bf {k}}}
\newcommand{\tk}{\tilde{k}}
\newcommand{\vtk}{\tilde{{\bf k}}}
\newcommand{\tK}{{\widetilde{K}}}
\newcommand{\vl}{{\bf {l}}}
\newcommand{\tl}{\tilde{l}}
\newcommand{\vtl}{\tilde{{\bf l}}}
\newcommand{\tL}{{\widetilde{L}}}
\newcommand{\cJ}{{\cal J}}
\newcommand{\cK}{{\cal K}}
\newcommand{\cL}{{\cal L}}
\newcommand{\ctJ}{\widetilde{{\cal J}}}
\newcommand{\ctK}{\widetilde{{\cal K}}}
\newcommand{\ctL}{\widetilde{{\cal L}}}
\newcommand{\m}{{\vec m}}
\newcommand{\hm}{{\hat m}}
\newcommand{\M}{{\vec M}}
\newcommand{\n}{{\vec n}}

\newcommand{\hn}{{\hat n}}
\newcommand{\vP}{{\bf \Pi}}
\newcommand{\vp}{{\bf p}}
\newcommand{\vtp}{\tilde{\bf p}}
\newcommand{\vR}{{\bf R}}
\newcommand{\hR}{{\hat R}}
\newcommand{\hvR}{\hat {\bf R}}
\newcommand{\tR}{\widetilde{R}}

\newcommand{\Si}{{\Sigma}}
\newcommand{\vSi}{{\bf {\Sigma}}}
\newcommand{\tU}{\widetilde {U}}
\newcommand{\tV}{\widetilde {V}}
\newcommand{\tW}{\tilde {W}}
\newcommand{\X}{{\vec X}}
\newcommand{\pX}{{\vec X}'}
\newcommand{\dX}{\dot{\vec X}}
\newcommand{\Y}{{\vec Y}}
\newcommand{\pY}{{\vec Y}'}
\newcommand{\dY}{\dot{\vec Y}}
\newcommand{\vZ}{{\bf Z}}
\newcommand{\tZ}{\widetilde{Z}}
\newcommand{\pZ}{{\bf Z}'}
\newcommand{\dZ}{\dot{\bf Z}}
\newcommand{\Za}{{Z}^1}
\newcommand{\Zb}{\widetilde{Z}^1}
\newcommand{\Zc}{{Z}^2}
\newcommand{\Zd}{\widetilde{Z}^2}
\begin{document}
\title {Abelian Decomposition of General Relativity}
\author{Y. M. Cho}
\email{ymcho@unist.ac.kr}
\author{S. H. Oh}
\affiliation{School of Electrical and Computer Engineering \\
Ulsan National University of Science and Technology, Ulsan 689-798 \\
and\\
School of Physics and Astronomy  \\
College of Natural Sciences\\ 
Seoul National University, Seoul 151-742, Korea}
\author{Sang-Woo Kim}
\affiliation{School of Physics and Astronomy, College of Natural Sciences\\ 
Seoul National University, Seoul 151-742, Korea}  
\begin{abstract}
~~~~~Based on the view that Einstein's theory can be interpreted 
as a gauge theory of Lorentz group, we decompose the gravitational
connection (the gauge potential of Lorentz group) $\vGm_\mu$ into
the restricted connection made of the potential of the maximal
Abelian subgroup $H$ of Lorentz group $G$ and the valence
connection made of $G/H$ part of the potential which transforms
covariantly under Lorentz gauge transformation. With this decomposition we show
that the Einstein's theory can be decomposed into the restricted
part made of the restricted connection which has the full Lorentz
gauge invariance and the valence part made of the valence
connection which plays the role of gravitational source of the
restricted gravity. We show that there are two different Abelian
decomposition of Einstein's theory, the light-like (or null) 
decomposition and the non light-like (or non-null) 
decomposition, because Lorentz group has two
maximal Abelian subgroups. In this decomposition the role of the
metric $g_\mn$ is replaced by a four-index metric
tensor $\vg_\mn$ which transforms covariantly under the Lorentz
group, and the metric-compatibility condition $\nabla_\alpha
g_\mn=0$ of the connection is replaced by the gauge and generally covariant
condition ${\mathscr D}_\mu \vg^\mn=0$. The decomposition 
shows the existence of a restricted theory of gravitation which has 
the full general invariance but is much simpler and has less physical 
degrees of freedom than Einstein's theory. Moreover, it tells that 
the restricted gravity can be written as an Abelian gauge theory, 
which implies that the graviton can be described by a massless spin-one 
field.
\end{abstract}

\maketitle

\section{Introduction}

Einstein's theory of gravitation and the gauge theory of electroweak and 
strong interactions are two fundamental ingredients of theoretical physics 
which describe all known interactions of nature.
But they are closely related to each other. The gauge
theory can be viewed as a part of Einstein's theory originating from
the extrinsic curvature of higher-dimensional unified space \cite{kal,jmp75}.
It is well-known that the (4+n)-dimensional
unified space made of the 4-dimensional
space-time and an n-dimensional internal space, the
(4+n)-dimensional Einstein's theory reproduces the gauge theory when
the internal space has an n-dimensional isometry $G$.
In fact the (4+n)-dimensional Einstein's theory provides a natural
unification of gauge theory with gravitation, which is known as the
Kaluza-Klein miracle \cite{jmp75,prd75}.

Conversely Einstein's theory itself can be understood as a
gauge theory, because the general invariance of Einstein's theory
can be viewed as a gauge invariance \cite{uti,kib}. One can view it as 
a gauge theory of 4-dimensional translation group, because the local
4-dimensional translation can be identified as the general
coordinate transformation. In this case
one can identify the gauge potential of the translation group
as the (non-trivial part of the) tetrad \cite{kib,prd76a}.
Or, one can view it as a gauge theory of Lorentz
group (or Poincare group in general), because
the Lorentz gauge transformation can also be
interpreted as the general coordinate
transformation. In this case one can identify
the gauge potential of Lorentz group as the spin
connection \cite{prd76b,hehl}. This confirms that
the two theories are closely related.

During the last few decades our understanding of non-Abelian gauge
theory has been extended very much. By now it has been well known
that the non-Abelian gauge theory allows the Abelian
decomposition \cite{prd80,prl81}. The non-Abelian gauge potential 
can be decomposed into the restricted potential of the maximal 
Abelian subgroup $H$ of the gauge group $G$
which has an electric-magnetic duality and the valence
potential of $G/H$ which transforms covariantly under $G$.
A remarkable feature of this decomposition is that
it is gauge independent. As importantly, the
restricted potential has the full non-Abelian gauge degrees of
freedom, in particular the topological degrees of the gauge group
$G$, in spite of the fact that it consists of only the Abelian
degrees of the maximal Abelian subgroup $H$. This means
that we can construct a restricted gauge theory, a non-Abelian
gauge theory made of only the restricted potential which has much
less physical degrees of freedom, which nevertheless has
the full gauge invariance. Moreover, we can recover the
full non-Abelian gauge theory simply by adding the valence part.
This tells that the non-Abelian gauge theory can be interpreted
as a restricted gauge theory which has the valence potential
as the gauge covariant source \cite{prd80,prl81}. The importance of
this decomposition is that the restricted part
plays a crucial role in non-Abelian dynamics, in particular in the
confinement mechanism in QCD \cite{fadd,prd02,jhep04,kondo}.

The main purpose of this paper is to discuss a similar Abelian
decomposition of Einstein's theory. Regarding the theory
as a gauge theory of Lorentz group and applying the Abelian 
decomposition to the gauge potential of Lorentz group,
we first show that we can decompose the gravitational connection
to the restricted connection and the valence connection.
With this we decompose the Einstein's theory into
the restricted part made of the restricted connection
and the valence part made of the gauge covariant valence connection. 
We show that Einstein's theory allows two different Abelian decompositions,
light-like decomposition and non light-like decomposition.
This is because the Lorentz group has two maximal Abelian subgroups.
With the Abelian decomposition we finally show that the restricted 
gravity can be interpreted as an Abelian gauge theory.  
{\it Our analysis tells that the Einstein's theory can be viewed as
a restricted theory of gravitation which has the gauge covariant
valence connection as the gravitational source. More importantly
our analysis implies that the graviton can be described by 
a massless spin-one gauge potential.}  

To decompose the Einstein's theory we introduce the gauge covariant 
metric $\vg_\mn$, an antisymmetric $(0,2)$-tensor in space-time 
$g_\mn^{~~ab}$ which forms an adjoint representation of Lorentz group, 
and show that the metric-compatibility condition of the gravitational 
connection $\nabla_\alpha g_\mn=0$ is transformed to the gauge covariant 
(Lorentz covariant) and generally covariant condition ${\mathscr D}_\mu \vg^\mn=0$ 
which assures the invariance of $\vg_\mn$ under the parallel transport along 
the $\pro_\mu$-direction.

Of course, Einstein's theory as a gauge theory of Lorentz group
is different from the ordinary non-Abelian gauge theory.
In gauge theory the fundamental field is the gauge potential,
but in Einstein's theory the fundamental field is the metric.
And in the gauge formulation the gauge potential of Lorentz group
corresponds to the gravitational connection, not the metric.
Also, in gauge theory the Yang-Mills Lagrangian is quadratic in
field strength. But in gravitation the
Einstein-Hilbert Lagrangian is made of the scalar curvature,
which is linear in field strength \cite{prd76b}.
Nevertheless we can still make the Abelian decomposition of the
gravitational connection, and express the Einstein-Hilbert Lagrangian
in terms of the restricted connection and the valence connection.
With this we can separate the restricted part of gravitation
from the Einstein's theory, and show that the theory
can be interpreted as a restricted theory of gravity
which has the valence connection as the gravitational source.

The paper is organized as follows. In Section II
we review the prototype Abelian decomposition, the $U(1)$
decomposition of $SU(2)$ gauge theory, as an example to help
us to understand the Abelian decomposition of
Einstein's theory. In Section III we show how to decompose
the gravitational connection to the Abelian part and the
valence part. We show that there are two different ways
of Abelian decomposition, because the Lorentz group
has two maximal Abelian subgroups. In Section IV introduce
the concept of the Lorentz covariant metric tensor, and show
how to decompose the Einstein's theory to the restricted
part and the valence part. We discuss two different Abelian
decompositions of Einstein's theory separately. In section V
we introduce two restricted gravities based on two Abelian decompositions, 
and show that they can be described by an Abelian gauge theory. 
In particular we argue that the graviton can be described by a massless 
spin-one gauge field. Finally in Section VI we discuss
the physical implications of our results.

\section{Abelian decomposition of SU(2): A Review}

To understand how the Abelian decomposition works in Einstein's theory, 
it is important to understand the Abelian decomposition $SU(2)$ gauge theory 
for two reasons.  First, it is the simplest non-Abelian gauge theory in which 
we can demonstrate the Abelian decomposition. But more importantly, it is 
the rotation subgroup of Lorentz group, so that the Abelian decomposition of 
$SU(2)$ directly applies to the Abelian decomposition of Einstein's theory. 
For these reasons we review the Abelian decomposition $SU(2)$ gauge theory 
first \cite{prd80,prl81}. 

Let $\hn$ be an arbitrary isotriplet unit vector field of $SU(2)$, and 
identify the maximal Abelian subgroup to be the $U(1)$ subgroup which 
leaves $\hn$ invariant. Clearly $\hn$ selects
the ``Abelian'' direction (i.e., the color charge direction)
at each space-time point, and the the Abelian magnetic isometry
can be described by the following constraint equation
\bea
D_\mu \hn = \pro_\mu \hn + g \A_\mu \times \hn = 0.
~~~(\hn^2=1)
\eea
This has the unique solution for $\A_\mu$ which defines
the restricted potential $\hA_\mu$ which leaves $\hn$ invariant 
under the parallel transport,
\bea
&\hA_\mu=A_\mu \hn - \oneg \hn\times\pro_\mu\hn,
\eea
where $A_\mu = \hn\cdot \A_\mu$ is the ``electric'' potential.
This process of selecting the restricted potential is called the
Abelian projection \cite{prd80,prl81}.

With the Abelian projection we can retrieve the full gauge
potential by adding the gauge covariant valence potential $\X_\mu$
to the restricted potential,
\bea
& \A_\mu =A_\mu \hn - \oneg \hn\times\pro_\mu\hn+\X_\mu
= \hA_\mu + \X_\mu, \nn\\
&(\hn^2 =1,~~~ \hn\cdot\X_\mu=0).
\label{su2dec}
\eea
This is the Abelian decomposition which decomposes
the gauge potential into the restricted potential
$\hA_\mu$ and the valence potential $\X_\mu$ \cite{prd80,prl81}.

Let $\vec \alpha$ is an infinitesimal gauge parameter.
Under the infinitesimal gauge transformation
\bea
&\delta \hn = - \vec \alpha \times \hn,~~~~~~ \delta \vec A_\mu =
\oneg D_\mu \vec \alpha,
\label{su2gt}
\eea
one has
\bea
&\delta A_\mu = \oneg \hn \cdot \pro_\mu \vec \alpha,~~~~ \delta
\hA_\mu = \oneg \hD_\mu \vec \alpha, \nn\\
&\delta \X_\mu = - \vec \alpha \times \X_\mu.
\eea
This shows that $\hA_\mu$ by itself describes an $SU(2)$ connection
which enjoys the full $SU(2)$ gauge degrees of freedom. Furthermore
$\X_\mu$ transforms covariantly under the gauge transformation. Most
importantly, the decomposition is gauge-independent. Once the color
direction $\hn$ is selected, the decomposition follows independent
of the choice of a gauge. This decomposition was first introduced
long time ago in an attempt to demonstrate the monopole condensation
in QCD \cite{prd80,prl81}. But recently the importance of the
decomposition in the non-Abelian dynamics has been emphasized by many 
authors \cite{fadd,kondo}. 

In particular, recently the Abelian decomposition 
has been successfully used in the lattice calculation of QCD to demonstrate 
the monopole condensation and color confinement in a gauge independent 
way \cite{kato}. A critical defect of the conventional lattice calculations 
is that the calculation is gauge dependent, because one has to choose 
a gauge (so-called the maximally Abelian gauge) to perform the calculation. 
With the Abelian decomposition, however, one does not have to choose a gauge 
to perform the calculation. So the recent calculation was able to demonstrate 
that the monopole condensation in QCD is a gauge independent phenomenon. 

To understand the physical meaning of our decomposition notice that
the restricted potential $\hA_\mu$ actually has a dual structure.
Indeed the field strength made of the restricted potential is
decomposed as
\bea
&\hF_\mn=\pro_\mu \hA_\nu-\pro_\nu \hA_\mu+ g
\hA_\mu \times \hA_\nu =(F_\mn+ H_\mn)\hn, \nn\\
&F_\mn=\pro_\mu A_\nu-\pro_\nu A_\mu, \nn\\
&H_\mn=-\oneg \hn \cdot(\pro_\mu\hn\times\pro_\nu\hn) =\pro_\mu
\tC_\nu-\pro_\nu \tC_\mu,
\eea
where $\tC_\mu$ is the ``magnetic'' potential \cite{prd80,prl81}.
Notice that we can always introduce the magnetic potential (at least
locally section-wise), because $H_\mn$ forms a closed two-form
\bea
\pro_\mu H_\mn^d = 0 ~~~~~~~ ( H_\mn^d = \dfrac12
\epsilon_{\mn\rho\sigma} H_{\rho\sigma} ).
\eea
This allows us to identify the non-Abelian magnetic potential by
\bea
\C_\mu= -\oneg \hn \times \pro_\mu\hn ,
\label{mp}
\eea
in terms of which the magnetic field strength is expressed as
\bea
&\vec H_\mn=\pro_\mu \C_\nu-\pro_\nu \C_\mu+ g
\C_\mu \times \C_\nu  \nn\\
&=-g \C_\mu \times \C_\nu = -\oneg \pro_\mu\hn\times\pro_\nu\hn
=H_\mn\hn.
\eea
As importantly $\hA_\mu$, as an $SU(2)$ potential, retains all the
essential topological characteristics of the original non-Abelian
potential. This is because the topological field $\hn$ naturally
represents the non-Abelian topology $\pi_2(S^2)$ which describes the
mapping from an $S^2$ in 3-dimensional space $R^3$ to the coset space
$SU(2)/U(1)$, and $\pi_3(S^3)\simeq\pi_3(S^2)$ which describes the
mapping from the compactified 3-dimensional space
$S^3$ to the group space $S^3$.
Clearly the isolated singularities of $\hn$ defines $\pi_2(S^2)$
which describes the non-Abelian monopoles.  Indeed $\C_\mu$ with
$\hn=\hat r$ describes precisely the Wu-Yang
monopole \cite{prd80,prl80}. This is why we call $\C_\mu$
the magnetic potential.
Besides, with the $S^3$ compactification of $R^3$, $\hn$
characterizes the Hopf invariant $\pi_3(S^2)\simeq\pi_3(S^3)$ which
describes the topologically distinct vacua \cite{bpst,baal,plb06}.

With (\ref{su2dec}) we have
\bea
\F_\mn=\hF_\mn + \hD_\mu \X_\nu - \hD_\nu \X_\mu + g\X_\mu \times
\X_\nu,
\eea
so that the Yang-Mills Lagrangian is expressed as
\bea
&{\cal L}=-\dfrac{1}{4} \F^2_\mn =-\dfrac{1}{4} \hF_\mn^2
-\dfrac{1}{4} ( \hD_\mu \X_\nu -\hD_\nu \X_\mu)^2 \nn\\
&-\dfrac{g}{2} \hF_\mn \cdot (\X_\mu \times \X_\nu)
- \dfrac{g^2}{4} (\X_\mu \times \X_\nu)^2 \nn\\
&+ \lambda(\hn^2 -1) + \lambda_\mu \hn \cdot \X_\mu,
\label{su2lag}
\eea
where $\lambda$ and $\lambda_\mu$ are the Lagrangian multipliers.
From the Lagrangian we have
\bea
&\delta A_\nu:~\pro_\mu (F_\mn+H_\mn+X_\mn) \nn\\
&= -g \hn \cdot \{ \X_\mu \times (\hD_\mu \X_\nu - \hD_\nu \X_\mu) \}, \nn\\
&\delta \X_\nu:~\hD_\mu ( \hD_\mu \X_\nu - \hD_\nu \X_\mu ) \nn\\
&=g (F_\mn+H_\mn+X_\mn) \hn \times \X_\mu, \nn\\
&X_\mn = g \hn \cdot ( \X_\mu \times \X_\nu ).
\label{su2eq}
\eea
Notice that here $\hn$ has no equation of motion even though the
Lagrangian contains it explicitly. This is because it represents
a topological degrees of freedom, not a
local degrees of freedom \cite{prd80,prl81}.
From this we conclude that the non-Abelian gauge
theory can be viewed as a restricted gauge theory made of the
restricted potential, which has an additional colored source made of
the valence gluon.

Obviously the Lagrangian (\ref{su2lag}) is invariant under the
active gauge transformation (\ref{su2gt}). But notice that the
decomposition introduces another gauge symmetry that we call the
passive gauge transformation \cite{prd02,jhep04},
\bea
\delta \hn = 0, ~~~~~~~\delta \A_\mu = \oneg D_\mu \vec
\alpha,\label{su2gt2}
\eea
under which we have
\bea
&\delta A_\mu = \oneg \hn \cdot D_\mu \vec \alpha, ~~~~~~~\delta
\hA_\mu = \oneg (\hn \cdot D_\mu \vec \alpha) \hn, \nn\\
&\delta \X_\mu = \oneg \{ D_\mu \vec \alpha -(\hn \cdot D_\mu \vec
\alpha) \hn \}.
\eea
This is because, for a given $\A_\mu$, one can have infinitely many
different decomposition of (\ref{su2dec}), with different $\hA_\mu$
and $\X_\mu$ choosing different $\hn$. Equivalently, for a fixed
$\hn$, one can have infinitely many different $\A_\mu$ which are
gauge-equivalent to each other. So our decomposition automatically
induce another type of gauge invariance which comes from different
choices of decomposition. This extra gauge invariance plays a
crucial role in quantizing the theory \cite{prl81}.

An important advantage of the decomposition (\ref{su2dec})
is that it can actually ``Abelianize'' (or more precisely ``dualize'')
the non-Abelian dynamics, without any gauge fixing \cite{prd80,prd02}.
To see this let $(\hn_1,\hn_2,\hn_3=\hn)$ be
a right-handed orthonormal basis and let
\bea
&\X_\mu =X^1_\mu ~\hn_1 + X^2_\mu ~\hn_2, \nn\\
&(X^1_\mu = \hn_1 \cdot \X_\mu,~~~X^2_\mu = \hn_2 \cdot \X_\mu) \nn
\eea
and find
\bea
&\hD_\mu \X_\nu =\{\pro_\mu X^1_\nu-g(A_\mu+ \tC_\mu)X^2_\nu \} \hn_1 \nn\\
&+ \{\pro_\mu X^2_\nu+ g (A_\mu+ \tC_\mu)X^1_\nu \}\hn_2,
\eea
where now the magnetic potential $\tC_\mu$
can be written explicitly as
\bea
&\tC_\mu=-\dfrac1g \n_1 \cdot \pro_\mu \n_2,
\eea
up to the $U(1)$ gauge transformation which leaves $\hn$ invariant.
So with
\bea
& \bA_\mu= A_\mu + \tC_\mu,
~~~\bF_\mn = \pro_\mu \bA_\nu - \pro_\nu \bA_\mu, \nn\\
&X_\mu = \dfrac{1}{\sqrt{2}} ( X^1_\mu + i X^2_\mu ),
\eea
one could express the Lagrangian explicitly in terms of the dual
potential $B_\mu$ and the complex vector field $X_\mu$,
\bea
&{\cal L}=-\dfrac{1}{4} \bF_\mn^2 -\dfrac12 |\bD_\mu X_\nu-\bD_\nu
X_\mu|^2 + ig \bF_\mn X_\mu^* X_\nu \nn\\
&-\dfrac12 g^2 \{(X_\mu^*X_\mu)^2-(X_\mu^*)^2 (X_\nu)^2\},  \nn\\
&\bD_\mu = \pro_\mu + ig \bA_\mu. 
\label{su2lag2}
\eea
Clearly this describes an Abelian gauge theory coupled to the
charged vector field $X_\mu$. But the important point here is that
the Abelian potential $\bA_\mu$ is given by the sum of the electric
and magnetic potentials $A_\mu+ \tC_\mu$. In this form the equations
of motion (\ref{su2eq}) is re-expressed as
\bea
&\pro_\mu(\bF_\mn+X_\mn)
= i g X^*_\mu (\bD_\mu X_\nu - \bD_\nu X_\mu )  \nn\\
&- i g X_\mu (\bD_\mu X_\nu - \bD_\nu X_\mu )^* , \nn\\
&\bD_\mu(\bD_\mu X_\nu- \bD_\nu X_\mu)
=ig X_\mu (\bF_\mn+X_\mn),  \nn\\
&X_\mn = - i g ( X_\mu^* X_\nu - X_\nu^* X_\mu ).
\eea
This shows that one can indeed Abelianize the non-Abelian theory
with our decomposition. The remarkable change in this ``Abelian''
formulation is that here the topological field $\hn$ is replaced by
the magnetic potential $\tC_\mu$.

\section{Abelian Decomposition of Gravitational Connection}

We can apply the above Abelian decomposition to Einstein's theory,
regarding Einstein's theory as a gauge theory of Lorentz group.
To do this we introduce a coordinate
basis
\bea
[\pro_\mu,~\pro_\nu] = 0, ~~~(\mu,\nu =t,x,y,z) \nn
\eea
and an orthonormal basis
\bea
&[\xi_a,~\xi_b] = f_{ab}^{~~c} \xi_c. ~~~(a,b= 0,1,2,3) \nn \\
&\xi_a=e_a^{~\mu} \pro_\mu,~~~~\pro_\mu=e_\mu^{~a} \xi_a,
\eea
where $e_\mu^{~a}$ and $e_a^{~\mu}$ are the tetrad and inverse tetrad.
Let $J_{ab}=-J_{ba}$ be the generators of Lorentz group,
\bea
&[J_{ab}, ~J_{cd}] = \eta_{ac} J_{bd}-\eta_{bc} J_{ad}
+\eta_{bd} J_{ac}-\eta_{ad} J_{bc} \nn\\
&=f_{ab,cd}^{~~~~~~mn}~J_{mn}, \nn\\
&f_{ab,cd}^{~~~~~~mn}=\eta_{ac} \delta_b^{~[m} \delta_d^{~n]}
-\eta_{bc} \delta_a^{~[m} \delta_d^{~n]} \nn\\
&+\eta_{bd} \delta_a^{~[m} \delta_c^{~n]} -\eta_{ad} \delta_b^{~[m}
\delta_c^{~n]},
\label{lgcr}
\eea
where $\eta_{ab}=diag~(-1,1,1,1)$ is the Minkowski metric. Clearly
$J_{ab}$ has the following 4-dimensional matrix representation
\bea
&(J_{ab})_{c}^{~d}=-\eta_{ac}\delta_{b}^{~d}
+\eta_{bc}\delta_{a}^{~d},
\eea
so that under the infinitesimal gauge transformation we have
\bea
&\delta~e_\mu^c=(\eta_{ad}\delta_{b}^{~c}
-\eta_{bd}\delta_{a}^{~c})~\alpha^{ab}~e_\mu^d,
\eea
where $\alpha^{ab}(=-\alpha^{ba})$ is an infinitesimal gauge parameter of
the Lorentz group. Instead of $(ab,cd,...)$ we can use
the index $(A,B,...)=(1,2,3,4,5,6)=(23,31,12,01,02,03)$, and write
\bea
[J_A, ~J_B] ~~=~~ f_{AB}^{~~~C} ~J_C~.  \nn
\eea
Moreover, with
\bea
&L_{1,2,3}=J_{1,2,3}=J_{23,31,12} \nn\\
&K_{1,2,3}=J_{4,5,6}=J_{01,02,03} \nn
\eea
the Lorentz algebra is written as
\bea
& [L_i, ~L_j] = \epsilon_{ijk} L_k,~~~~~[L_i, ~K_j] = \epsilon_{ijk} K_k, \nn \\
& [K_i, ~K_j] =-\epsilon_{ijk} L_k, ~~~(i,j,k= 1,2,3)
\eea
where $L_i$ and $K_i$ are the 3-dimensional rotation and
boost generators. Notice that the generators can be viewed
as the left-invariant basis vector fields
on the Lorentz group manifold which satisfy
the commutation relation.

As we have pointed out, we can regard Einstein's theory as a gauge
theory of Lorentz group. In this view the gravitational connection
$\Gamma_\mn^{~~\rho}$ (or more precisely the spin connection
$\omega_\mu^{~ab}$) corresponds to the gauge potential
$\Gamma_\mu^{~ab}$, and the curvature tensor
$R_\mn^{~~ab}$ corresponds to the gauge field strength
$F_\mn^{~~ab}$ of Lorentz group.
And to obtain the desired decomposition we have to decompose the
gauge potential $\Gamma_\mu^{~ab}$ first. Now, to apply the above $SU(2)$
decomposition to Lorentz group, we have to keep in mind that there
are notable differences between $SU(2)$ and Lorentz group. First,
the Lorentz group is non-compact, so that the invariant metric is
indefinite. Secondly, the Lorentz group has the well-known invariant
tensor $\epsilon_{abcd}$ which allows the dual transformation.
Thirdly, the Lorentz group has rank two, so that it has two
commuting Abelian subgroups and two Casimir invariants.
Finally, the Lorentz group has two different maximal
Abelian subgroups $A_2$ and $B_2$ \cite{jmp79}.
These differences make the
decomposition more complicated.

The invariant metric $\delta_{AB}$ of Lorentz group
is given by
\bea
&\delta_{AB}= -\dfrac{1}{4} f_{AC}^{~~D}f_{BD}^{~~C} \nn\\
&=diag~(+1,+1,+1,-1,-1,-1).
\label{inme}
\eea
Let $p^{ab}~(p^{ab}=-p^{ba})$ (or $p^A$) be a gauge covariant sextet
vector which forms an adjoint representation of Lorentz group,
\bea
\delta~p^{cd}= - \dfrac12 f_{ab,mn}^{~~~~~~cd} \alpha^{ab}~p^{mn}.
\eea
Clearly $p^{ab}$ can be understood as an anti-symmetric tensor in
4-dimensional Minkowski space which can be expressed by two
3-dimensional vectors $\m$ and $\e$, which transform exactly like
the magnetic and electric components of an electromagnetic tensor
under the 4-dimensional Lorentz transformation. And we denote $p^{ab}$ by $\vp$, 
\bea
&\vp = \dfrac12 p_{ab} \vI^{ab}=\left( \begin{array}{c} \m \\
\e \end{array} \right),
~~p^{ab}=\vp \cdot \vI^{ab}=\dfrac12 p^{mn} I_{mn}^{~~~ab}, \nn\\
&\vI^{ab}=\left( \begin{array}{c} {\hat m}^{ab} \\
{\hat e}^{ab} \end{array} \right), \nn\\
&{\hat m}_i^{~ab}=\epsilon_{0i}^{~~ab}, 
~~~{\hat e}_i^{~ab}=\big(\delta_0^{~a} \delta_i^{~b} -\delta_0^{~b}
\delta_i^{~a} \big), \nn\\
&I_{mn}^{~~~ab}=\big(\delta_m^{~a} \delta_n^{~b} -\delta_m^{~b}
\delta_n^{~a} \big) =-(J_{mn})^{ab}.
\label{iddef}
\eea
where $m_i=\epsilon_{ijk}p^{jk}/2~(i,j,k=1,2,3)$ is the magnetic
(or rotation) part and $e_i=p^{0i}$  is the electric (or boost) part of
$\vp$. From the invariant metric (\ref{inme}) we have
\bea
\vp^2= \dfrac 12 p_{ab} p^{ab}= \m^2-\e^2,
\eea
so that the invariant length can be positive, zero, or negative.
This, of course, is due to the fact that the invariant metric
(\ref{inme}) is indefinite.

The Lorentz group has another important invariant tensor
$\epsilon_{AB}$ which comes from the totally anti-symmetric
invariant tensor $\epsilon_{abcd}$,
\bea
\epsilon_{AB}=\epsilon_{ab,cd}=\epsilon_{abcd}.
\eea
This tells that any adjoint representation of Lorentz group has its
dual partner. In particular, $\vp$ has the dual vector $\vtp$
defined by $\widetilde p^{ab}=\epsilon^{abcd} p_{cd}/2$. With
(\ref{iddef}) we have (with $\epsilon_{0123}=+1$)
\bea
&\vtp=\left( \begin{array}{c} \e \\ -\m
\end{array} \right),
~~~~~\tilde{\vtp}=-\vp, \nn\\
&\vtp^2=\e^2-\m^2=-\vp^2,  \nn\\
&\vp \cdot \vtp= \dfrac{1}{4} \epsilon_{abcd} p^{ab} p^{cd}
=2 \m \cdot \e.
\eea
Moreover, we have
\bea
[p, ~\widetilde p~]=0,~~~~~\vp \times \vtp=0.
\label{cl}
\eea
This tells that any two vectors which are dual to each other are
always commuting. Finally we have the following vector operations,
\bea
&{\vp \cdot \vp'}= \m \cdot \m'-\e \cdot \e', \nn\\
&{\vp \cdot \vtp'}= \m \cdot \e'+\e \cdot \m'={\vtp \cdot \vp'}, \nn\\
&\vp \times \vp'=\left(\begin{array}{c} \m \times \m'-\e \times \e' \\
\m \times \e'+\e \times \m'  \end{array} \right)
=- \vtp \times \vtp', \nn\\
&\vp \times \vtp'= \left(\begin{array}{c} \m \times \e' +\e
\times\m' \\ -\m \times \m' +\e \times \e'
\end{array} \right) = \vtp \times \vp', \nn\\
&\widetilde{\vp \times \vp'}=\vp \times \vtp'=\vtp \times \vp', \nn\\
&\vp_1 \cdot (\vp_2 \times \vp_3)=\vp_2 \cdot (\vp_3 \times \vp_1)
=\vp_3 \cdot (\vp_1 \times \vp_2),  \nn\\
&\vp_1 \times (\vp_2 \times \vp_3)=[\vp_2~(\vp_1 \cdot \vp_3)-\vp_3~(\vp_1 \cdot \vp_2)]  \nn\\
&-[\vtp_2~(\vp_1 \cdot \vtp_3)-\vtp_3~(\vp_1 \cdot \vtp_2)],  
\eea
so that we can always reduce the operations of 6-dimensional vectors
of Lorentz group to the operations of 3-dimensional vectors.

Let $(\hn_1,\hn_2,\hn_3=\hn)$ be a $3$-dimensional unit vectors
($\hn_i^2=1$) which form a right-handed orthonormal
basis with $\hn_1 \times \hn_2=\hn_3$, and let
\bea
&\vl_i= \left( \begin{array}{c} \hn_i \\
0 \end{array} \right),
~~~\vk_i= \left( \begin{array}{c} 0 \\
\hn_i  \end{array} \right)= -\vtl_i. \label{lbasis}
\eea
Clearly we have
\bea
&\vl_i \cdot \vl_j=\delta_{ij}, ~~~\vl_i \cdot \vk_j=0,
~~~\vk_i\cdot \vk_j=-\delta_{ij},  \nn\\
&\vl_i \times \vl_j= \epsilon_{ijk} \vl_k,~~~~~\vl_i \times \vk_j= \epsilon_{ijk} \vk_k, \nn\\
&\vk_i \times \vk_j= -\epsilon_{ijk} \vl_k
\eea
so that $(\vl_i,\vk_i)$, or equivalently $(\vl_i,\vtl_i)$,
forms an orthonormal basis of
the adjoint representation of Lorentz group.

To make the desired Abelian decomposition we have to choose the gauge
covariant sextet vector fields which form adjoint representation of
Lorentz group which describe the desired magnetic isometry. To see
what types of isometry is possible, it is important to
remember that Lorentz group has two $2$-dimensional maximal Abelian 
subgroups, $A_2$ whose generators are made of $L_3$ and $K_3$ and
$B_2$ whose generators are made of $(L_1+K_2)/\sqrt 2$ and 
$(L_2-K_1)/\sqrt 2$ \cite{jmp79}. 

This tells that we have two possible
Abelian decompositions of the gravitational connection.
And in both cases the magnetic isometry is
described by two, not one, commuting sextet vector fields of Lorentz
group which are dual to each other. To see this let us denote one of
the isometry vector field by $\vp$ which satisfy the isometry
condition
\bea
D_\mu \vp = (\pro_\mu + \vGm_\mu \times) ~\vp=0, \label{ic}
\eea
where we have normalized the coupling constant to be the unit
(which one can always do without loss of generality).
Now, notice that the above condition automatically assures
\bea
D_\mu \vtp =(\pro_\mu + \vGm_\mu \times) ~\vtp=0, \label{dic}
\eea
because $\epsilon_{abcd}$ is an invariant tensor. This tells that
when $\vp$ is an isometry, $\vtp$ also becomes an isometry. To
verify this directly we decompose the gauge potential of Lorentz
group $\vGm_\mu$ into the 3-dimensional rotation and boost parts
$\A_\mu$ and $\B_\mu$, and let
\bea
\vGm_\mu= \left( \begin{array}{c} \A_\mu \\
\B_\mu \end{array} \right).
\eea
With this both (\ref{ic}) and (\ref{dic}) can be written as
\bea
&D_\mu \m= \B_\mu \times \e,~~~&D_\mu \e= -\B_\mu \times \m,
\label{ic1}
\eea
where now
\bea
D_\mu = \pro_\mu + \A_\mu \times. \nn
\eea
This confirms that (\ref{ic}) and (\ref{dic}) are actually identical
to each other, which tells that the magnetic isometry in Lorentz
group must be even-dimensional.

Since Lorentz group has two invariant tensors it has two Casimir
invariants. And it is useful to characterize the isometry by two
Casimir invariants. Let the isometry be described by $\vp$ and
$\vtp$. It has two Casimir invariants $\alpha$ and $\beta$,
\bea
&\alpha ~=~ {\vp \cdot \vp} ~=~ \m^2-\e^2, \nn\\
&\beta ~=~ {\vp \cdot \vtp} ~=~2 \m \cdot \e.
\eea
But the Casimir invariants $(\alpha,\beta)$
depends on the choice of the isometry vectors.
To see this consider $\vp'$ and $\vtp'$ given by
a linear combination of $\vp$ and $\vtp$,
\bea
&\vp'= a\vp+b\vtp,
~~~~~\vtp'=a\vtp -b\vp.
\eea
Clearly we have
\bea
&D_\mu \vp'=0,~~~~~D_\mu \vtp'=0,
\eea
so that they can also be viewed to describe the same isometry.
But their Casimir invariants $(\alpha',\beta')$
are given by
\bea
&\alpha'= (a^2-b^2)\alpha+2ab\beta, \nn\\
&\beta'=(a^2-b^2)\beta-2ab\alpha.
\eea
And with
\bea
&a=\sqrt{\dfrac{(\alpha^2+\beta^2)^{1/2} \pm \alpha}{2(\alpha^2+\beta^2)}},  \nn\\
&b=\pm \dfrac{|\beta|}{\beta} 
\sqrt{\dfrac{(\alpha^2+\beta^2)^{1/2} \mp \alpha}{2(\alpha^2+\beta^2)}}, \nn
\eea
we can always make
\bea
&\alpha'= \pm 1,~~~~~\beta'= 0,
\eea
unless $\alpha^2+\beta^2=0$. This tells that we can always choose
$\vp$ and $\vtp$ in such a way to make $(\alpha,\beta)$ to be
$(\pm 1,0)$ or $(0,0)$. Physically this means that the magnetic
isometry in Einstein's theory can be classified by the non light-like
(or space/time) isometry and the light-like (or null) isometry
whose Casimir invariants are denoted by $(\pm 1,0)$ and $(0,0)$,
respectively. We emphasize that once $\vp$ and $\vtp$ are
chosen, $(\alpha,\beta)$ are uniquely fixed. Now we discuss the
two isometries $A_2$ and $B_2$ separately.

\subsection{$A_2$ (Non Light-like) Isometry}

Let the maximal Abelian subgroup be $A_2$.
In this case the isometry is made of $L_3$ and $K_3$, and we have
two sextet vector fields which describes the isometry which are dual
to each other. Let $\vp$ and $\vtp$ be the two isometry vector
fields which correspond to $L_3$ and $K_3$. Clearly we can put
\bea
&\vp= f~\vl_3= f \left( \begin{array}{c} \hn \\
0  \end{array} \right),
~~~\vtp=f~\vtl_3= f \left( \begin{array}{c} 0 \\
-\hn  \end{array} \right),
\eea
where $f$ is an arbitrary function of space-time.
The Casimir invariants of the isometry vectors
are given by $(f^2,0)$.
But just as in $SU(2)$ gauge theory the isometry
condition (\ref{ic}) requires $f$ to be a constant,
because
\bea
\pro_\mu f^2=\pro_\mu \vp^2=D_\mu \vp^2=2\vp \cdot D_\mu \vp=0.
\eea
And we can always normalize $f=1$ without loss of generality.

So the $A_2$ isometry can always be written as
\bea
&\vl=\vl_3= \left( \begin{array}{c} \hn \\
0  \end{array} \right),
~~~~~\vtl=\vtl_3=\left( \begin{array}{c} 0 \\ -\hn
\end{array} \right), \nn\\
&{D_\mu \vl} =0,~~~~~{D_\mu \vtl}=0,
\label{a2ic}
\eea
whose Casimir invariants are fixed by $(1,0)$.
With this we find the restricted connection $\hvGm_\mu$
which satisfies the isometry condition
\bea
&\hvGm_\mu= A_\mu ~\vl - B_\mu ~\vtl
-\vl \times \pro_\mu \vl, \nn\\
&A_\mu = {\vl \cdot \vGm_\mu},
~~~B_\mu  = \vtl \cdot \vGm_\mu,
\label{a2rc}
\eea
where $A_\mu$ and $B_\mu$ are two Abelian connections
of $\vl$ and $\vtl$ components which are not restricted by
the isometry condition. At first glance this expression
appears strange, because one expects that $\vl$ and $\vtl$ should
contribute equally in the restricted connection since (\ref{ic}) and
(\ref{dic}) are identical. Actually they do contribute equally
because we have
\bea
&\vl \times \pro_\mu \vl=-\vtl \times \pro_\mu \vtl,
\eea
so that we can express the restricted connection as
\bea
&\hvGm_\mu= A_\mu ~\vl - B_\mu ~\vtl
-\dfrac12(\vl \times \pro_\mu \vl-\vtl \times \pro_\mu \vtl).
\eea
The restricted field strength $\hvR_\mn$ which describes
the restricted curvature tensor $\hR_\mn^{~~ab}$ is given by
\bea
&\hvR_\mn=\pro_\mu \hvGm_\nu-\pro_\nu \hvGm_\mu
+\hvGm_\mu \times \hvGm_\nu \nn\\
&= (A_\mn+ H_\mn) ~\vl - (B_\mn+\tH_\mn) ~\vtl, \nn\\
&A_\mn = \pro_\mu A_\nu - \pro_\nu A_\mu,
~~~B_\mn= \pro_\mu B_\nu - \pro_\nu B_\mu, \nn\\
&H_\mn = -\vl \cdot (\pro_\mu \vl \times \pro_\nu \vl), \nn\\
&\tH_\mn =-\vtl \cdot (\pro_\mu \vl \times \pro_\nu \vl)
=\vtl \cdot (\pro_\mu \vtl \times \pro_\nu \vtl)  \nn=0,  \nn\\
&\hR_\mn^{~~ab}=\hvR_\mn \cdot \vI^{ab} \nn\\
&= (A_\mn+ H_\mn) ~l^{ab} - B_\mn ~\tl^{ab}.
\label{a2rct}
\eea
Notice that $\tH_\mn$ vanishes.

In 3-dimensional notation the isometry condition (\ref{a2ic}) can be
written as
\bea
&\hat \vGm_\mu=\left( \begin{array}{c} \hat A_\mu \\
\hat B_\mu  \end{array} \right), \nn\\
&\hD_\mu \hn=0,~~~~~\hB_\mu \times \hn=0, \nn\\
&\hD_\mu=\pro_\mu+\hA_\mu \times.
\eea
From this we have
\bea
&\hat A_\mu=A_\mu \hn - \hn \times \pro_\mu \hn,
~~~\hat B_\mu=B_\mu \hn, \nn\\
&A_\mu=\hn \cdot \hat A_\mu,~~~B_\mu=\hn \cdot \hat B_\mu.
\label{a2rc3}
\eea
Moreover, with
\bea
\hvR_\mn=\left( \begin{array}{c} \hA_\mn \\
\hB_\mn  \end{array} \right), \label{rct}
\eea
we have
\bea
&\hA_\mn=\pro_\mu \hA_\nu-\pro_\nu \hA_\mu
+\hA_\mu \times \hA_\nu \nn\\
&=(A_\mn + H_\mn) \hn=\bar A_\mn \hn, \nn\\
&\hB_\mn=\pro_\mu \hB_\nu-\pro_\nu \hB_\mu
+\hA_\mu \times \hB_\nu-\hA_\nu \times \hB_\mu \nn\\
&=\hD_\mu \hB_\nu-\hD_\nu \hB_\mu =B_\mn ~\hn, \nn\\
&H_\mn=-\hn \cdot (\pro_\mu \hn \times \pro_\nu \hn)
=\pro_\mu \tC_\nu - \pro_\nu \tC_\mu, \nn\\
&\tC_\mu= -\hn_1 \cdot \pro_\mu \hn_2, \nn\\
&\bar A_\mn=\pro_\mu \bar A_\nu - \pro_\nu \bar A_\mu,
~~~~~\bar A_\mu= A_\mu +\tC_\mu.  
\eea
Notice that $\hA_\mu$ and $\hA_\mn$ are formally identical to the
restricted potential and restricted field strength of $SU(2)$ gauge
theory. In particular $H_\mn$ is identical to what we have in
Section II. This, together with $\tH_\mn =0$, tells that
the topology of this isometry is identical to that of the $SU(2)$
subgroup.

With this the full connection of Lorentz group is given by
\bea
&\vGm_\mu = \hvGm_\mu + \vZ_\mu, ~~~{\vl \cdot \vZ_\mu = \vtl \cdot
\vZ_\mu} = 0,
\eea
where $\vZ_\mu$ is the valence connection which transforms
covariantly under the Lorentz gauge transformation,
or equivalently under the general coordinate transformation.
The corresponding field strength $\vR_\mn$ which describes
the curvature tensor is written as
\bea
&\vR_\mn = \pro_\mu \vGm_\nu-\pro_\nu \vGm_\mu
+\vGm_\mu \times \vGm_\nu \nn\\
&=\hvR_\mn+\vZ_\mn, \nn\\
&\vZ_\mn=\hD_\mu \vZ_\nu - \hD_\nu \vZ_\mu
+ \vZ_\mu \times \vZ_\nu, \nn\\
&\hD_\mu=\pro_\mu + \hvGm_\mu \times, \label{dct}
\eea
where $\vZ_\mn$ is the valence part of the curvature tensor
which can further be decomposed to the kinetic part $\dZ_\mn$
and the potential part $\pZ_\mn$,
\bea
&\vZ_\mn=\dZ_\mn+\pZ_\mn, \nn\\
&\dZ_\mn=\hD_\mu \vZ_\nu - \hD_\nu \vZ_\mu,
~~~\pZ_\mn=\vZ_\mu \times \vZ_\nu.
\eea
Now with
\bea
&\vZ_\mu = \Za_\mu \vl_1 -\Zb_\mu \vtl_1
+\Zc_\mu \vl_2 -\Zd_\mu \vtl_2, \nn\\
&\Za_\mu =\vl_1 \cdot \vZ_\mu,~~~\Zb_\mu =\vtl_1 \cdot \vZ_\mu, \nn\\
&\Zc_\mu =\vl_2 \cdot \vZ_\mu,~~~\Zd_\mu =\vtl_2 \cdot \vZ_\mu,
\eea
we have
\bea
&\dZ_\mn= (\cD_\mu \Za_\nu-\cD_\nu \Za_\mu)\vl_1
-(\cD_\mu \Zb_\nu-\cD_\nu \Zb_\mu)\vtl_1 \nn\\
&+(\cD_\mu \Zc_\nu-\cD_\nu \Zc_\mu)\vl_2
-(\cD_\mu \Zd_\nu-\cD_\nu \Zd_\mu)\vtl_2, \nn\\
&\cD_\mu \Za_\nu = \pro_\mu \Za_\nu -\bA_\mu \Zc_\nu
+B_\mu \Zd_\nu,\nn\\
&\cD_\mu \Zb_\nu = \pro_\mu \Zb_\nu -\bA_\mu \Zd_\nu
-B_\mu \Zc_\nu,\nn\\
&\cD_\mu \Zc_\nu = \pro_\mu \Zc_\nu +\bA_\mu \Za_\nu
-B_\mu \Zb_\nu, \nn\\
&\cD_\mu \Zd_\nu = \pro_\mu \Zd_\nu +\bA_\mu \Zb_\nu
+B_\mu \Za_\nu, \nn\\
&\vl \cdot \dZ_\mn=\vtl \cdot \dZ_\mn=0.
\eea
Clearly $\bA_\mu$ is identical to the dual potential
we have introduced in Section II in $SU(2)$ gauge theory.
Moreover, we have
\bea
&\pZ_\mn=W_\mn \vl - \tW_\mn \vtl, \nn\\
& W_\mn = {\vl \cdot (\vZ_\mu \times \vZ_\nu)} \nn\\
&=\Za_\mu\Zc_\nu-\Za_\nu\Zc_\mu -(\Zb_\mu\Zd_\nu-\Zb_\nu\Zd_\mu), \nn\\
&\tW_\mn = \vtl \cdot (\vZ_\mu \times \vZ_\nu)  \nn\\
&=\Za_\mu\Zd_\nu-\Za_\nu\Zd_\mu +\Zb_\mu\Zc_\nu-\Zb_\nu\Zc_\mu.
\eea
With this we have the full curvature tensor
\bea
&\vR_\mn = (\bA_\mn+ W_\mn){\vl}-(B_\mn+ \tW_\mn){\vtl} \nn\\
&+ \hD_\mu \vZ_\nu - \hD_\nu \vZ_\mu \nn\\
&=(\cD_\mu \bA_\nu-\cD_\nu \bA_\mu){\vl}
-(\cD_\mu B_\nu-\cD_\nu B_\mu)\vtl \nn\\
&+(\cD_\mu \Za_\nu-\cD_\nu \Za_\mu)\vl_1
-(\cD_\mu \Zb_\nu-\cD_\nu \Zb_\mu)\vtl_1 \nn\\
&+(\cD_\mu \Zc_\nu-\cD_\nu \Zc_\mu)\vl_2
-(\cD_\mu \Zd_\nu-\cD_\nu \Zd_\mu)\vtl_2 \nn\\
&=R_\mn^1~\vl_1 -\tR_\mn^1~\vtl_1
+R_\mn^2~\vl_2 -\tR_\mn^2~\vtl_2 \nn\\
&+ R_\mn ~{\vl} - \tR_\mn ~\vtl,  \nn\\
&\cD_\mu \bA_\nu=\pro_\mu \bA_\nu+\Za_\mu\Zc_\nu-\Zb_\mu\Zd_\nu,  \nn\\
&\cD_\mu B_\nu=\pro_\mu B_\nu+\Za_\mu\Zd_\nu+\Zb_\mu\Zc_\nu, \nn\\
&R_\mn^1=\cD_\mu \Za_\nu-\cD_\nu \Za_\mu,
~~~\tR_\mn^1=\cD_\mu \Zb_\nu-\cD_\nu \Zb_\mu, \nn\\
&R_\mn^2=\cD_\mu \Zc_\nu-\cD_\nu \Zc_\mu,
~~~\tR_\mn^2=\cD_\mu \Zd_\nu-\cD_\nu \Zd_\mu,  \nn\\
&R_\mn=\cD_\mu \bA_\nu-\cD_\nu \bA_\mu
=A_\mn+ H_\mn+ W_\mn, \nn\\
&\tR_\mn=\cD_\mu B_\nu-\cD_\nu B_\mu=B_\mn+ \tW_\mn,
\label{a2ct1}
\eea
or equivalently
\bea
&R_\mn^{~~ab}=\vR_\mn \cdot \vI^{ab}  \nn\\
&=R_\mn^1~l_1^{ab}-\tR_\mn^1~\tl_1^{ab}
+R_\mn^2~l_2^{ab}-\tR_\mn^2~\tl_2^{ab}  \nn\\
&+R_\mn ~l^{ab} - \tR_\mn ~\tl^{ab}.
\label{a2ct2}
\eea
This is the $A_2$ decomposition of the curvature tensor.
The similarity between this decomposition and the
Abelian decomposition of $SU(2)$ is unmistakable.

To emphasize the similarity between this isometry and the
$U(1)$ isometry of $SU(2)$ we introduce the complex notation
\bea
&Z_\mu=\dfrac{1}{\sqrt2}(\Za_\mu+i\Zc_\mu),
~~~~ \tZ_\mu= \dfrac{1}{\sqrt2}(\Zb_\mu+i\Zd_\mu), \nn\\
&\vl_\pm=\dfrac{1}{\sqrt2}(\vl_1\pm i\vl_2),
~~~~\vtl_\pm=\dfrac{1}{\sqrt2}(\vtl_1\pm i\vtl_2),
\eea
and find
\bea
&\dZ_\mn= (\cD_\mu Z_\nu-\cD_\nu Z_\mu)^*~\vl_+
+(\cD_\mu Z_\nu-\cD_\nu Z_\mu)~\vl_-  \nn\\
&-(\cD_\mu \tZ_\nu-\cD_\nu \tZ_\mu)^*~\vtl_+
-(\cD_\mu \tZ_\nu-\cD_\nu \tZ_\mu)~\vtl_-, \nn\\
&\cD_\mu Z_\nu =(\pro_\mu +i \bA_\mu) Z_\nu
-i B_\mu \tZ_\nu=\bD_\mu Z_\nu-i B_\mu \tZ_\nu, \nn\\
&\cD_\mu \tZ_\nu =(\pro_\mu +i \bA_\mu) \tZ_\nu
+i B_\mu Z_\nu=\bD_\mu \tZ_\nu+i B_\mu Z_\nu, \nn\\
&\bD_\mu = \pro_\mu +i \bA_\mu.
\eea
Here $\bD_\mu$ is identical to the one we have in Section II.
Moreover, the potential part of $\vZ_\mn$ is given by
\bea
&\pZ_\mn= W_\mn\vl-\tW_\mn\vtl, \nn\\
&W_\mn = \Za_\mu\Zc_\nu-\Za_\nu\Zc_\mu
-(\Zb_\mu\Zd_\nu-\Zb_\nu\Zd_\mu) \nn\\
&=-i(Z^*_\mu Z_\nu-Z^*_\nu Z_\mu) 
+i(\tZ^*_\mu \tZ_\nu-\tZ^*_\nu \tZ_\mu), \nn\\
&\tW_\mn = \Za_\mu\Zd_\nu-\Za_\nu\Zd_\mu
+\Zb_\mu\Zc_\nu-\Zb_\nu\Zc_\mu \nn\\
&=-i(Z^*_\mu \tZ_\nu-Z^*_\nu \tZ_\mu) 
-i(\tZ^*_\mu Z_\nu-\tZ^*_\nu Z_\mu).
\eea
With this we have
\bea
&\vR_\mn = (\cD_\mu Z_\nu-\cD_\nu Z_\mu)^*~\vl_+
-(\cD_\mu \tZ_\nu-\cD_\nu \tZ_\mu)^*~\vtl_+  \nn\\
&+(\cD_\mu Z_\nu-\cD_\nu Z_\mu)~\vl_-
-(\cD_\mu \tZ_\nu-\cD_\nu \tZ_\mu)~\vtl_-  \nn\\
&+(\cD_\mu \bA_\nu-\cD_\nu \bA_\mu)~\vl
-(\cD_\mu B_\nu-\cD_\nu B_\mu)~\vtl,
\label{a2ct3}
\eea
or
\bea
&R_\mn^{~~ab}=(\cD_\mu Z_\nu-\cD_\nu Z_\mu)^*~l_+^{ab}
-(\cD_\mu \tZ_\nu-\cD_\nu \tZ_\mu)^*~\tl_+^{ab} \nn\\
&+(\cD_\mu Z_\nu-\cD_\nu Z_\mu)~l_-^{ab}
-(\cD_\mu \tZ_\nu-\cD_\nu \tZ_\mu)~\tl_-^{ab}  \nn\\
&+(\cD_\mu \bA_\nu-\cD_\nu \bA_\mu)~l^{ab}
-(\cD_\mu B_\nu-\cD_\nu B_\mu)~\tl^{ab}.
\label{a2ct4}
\eea
This should be compared with the $SU(2)$ decomposition.

In 3-dimensional notation we have
\bea
&\vZ_\mu=\left( \begin{array}{c} \X_\mu \\
\Y_\mu  \end{array} \right), \nn\\
&\X_\mu=Z_\mu^1~\hn_1+ Z_\mu^2~\hn_2,
~~~\Y_\mu=\tZ_\mu^1~\hn_1+ \tZ_\mu^2~\hn_2,  \nn\\
&\hn \cdot \X_\mu=0,~~~~~\hn \cdot \Y_\mu=0.
\label{3dvc}
\eea
Moreover, with
\bea
&\vZ_\mn=\left( \begin{array}{c} \X_\mn \\
\Y_\mn  \end{array} \right) =\left( \begin{array}{c} \dX_\mn+\pX_\mn \\
\dY_\mn+\pY_\mn  \end{array} \right),
\eea
we have
\bea
&\dX_\mn=\hD_\mu \X_\nu -\hD_\nu \X_\mu
-\B_\mu \times \Y_\nu+\B_\nu \times \Y_\mu \nn\\
&= R_\mn^1~\hn_1+ R_\mn^2~\hn_2, \nn\\
&\dY_\mn=\hD_\mu \Y_\nu -\hD_\nu \Y_\mu
+\B_\mu \times \X_\nu -\B_\nu \times \X_\mu \nn\\
&=\tR_\mn^1~\hn_1+ \tR_\mn^2~\hn_2, \nn\\
&\pX_\mn= \X_\mu \times \X_\nu-\Y_\mu \times \Y_\nu =W_\mn~\hn, \nn\\
&\pY_\mn= \X_\mu \times \Y_\nu+\Y_\mu \times \X_\nu =\tW_\mn~\hn.
\eea
Notice that the kinetic part and the potential part of $\vZ_\mn$ are
orthogonal to each other.
Finally, with
\bea
&\vR_\mn=\left( \begin{array}{c} \A_\mn \\
\B_\mn  \end{array} \right)=\left( \begin{array}{c} \hA_\mn+\X_\mn \\
\hB_\mn+\Y_\mn  \end{array} \right),
\label{fct}
\eea
we have
\bea
&\A_\mn=R_\mn ~\hn +\dX_\mn \nn\\
&=R_\mn^1~\hn_1+R_\mn^2~\hn_2+R_\mn ~\hn, \nn\\
&\B_\mn=\tR_\mn ~\hn +\dY_\mn \nn\\
&=\tR_\mn^1~\hn_1+\tR_\mn^2~\hn_2+\tR_\mn ~\hn.
\eea
This completes the $A_2$ decomposition of the gravitational
connection.

\subsection{$B_2$ (Light-like) Isometry}

This is when the isometry group is made of $(L_1+K_2)/\sqrt 2$ and
$(L_2-K_1)/\sqrt 2$.
Let $\vp$ and $\vtp$ be the two isometry vector fields which
correspond to $(L_1+K_2)/\sqrt 2$ and $(L_2-K_1)/\sqrt 2$ which are
dual to each other. In this case we can write
\bea
&{\vp}= f\Big(\dfrac{\vl_1+ \vk_2}{\sqrt 2}\Big)
=\dfrac{f}{\sqrt 2} \left( \begin{array}{c} \hn_1 \\
 \hn_2  \end{array} \right), \nn\\
&{\vtp}=f\Big(\dfrac{ \vl_2-\vk_1}{\sqrt 2}\Big)
=\dfrac{f}{\sqrt 2} \left( \begin{array}{c}  \hn_2 \\
- \hn_1  \end{array} \right).
\eea
But notice that the Casimir invariants ($\alpha,\beta$)
of the isometry vectors are given by
($0,0$) independent of $f$. Moreover, here
(unlike the $A_2$ case) the isometry
condition does not restrict $f$ at all, because we have
$\vp^2=0$ independent of $f$. So the $B_2$ isometry vectors contain
an arbitrary scalar function $f(x)$.

Let us put $f=e^{\lambda}$ and express the $B_2$ isometry by
\bea
&{\vj}= \dfrac{e^\lambda}{\sqrt2} (\vl_1+ \vk_2)
=\dfrac{e^\lambda}{\sqrt2} \left(\begin{array}{c} \hn_1 \\
\hn_2  \end{array} \right),  \nn\\
&{\vtj}= \dfrac{e^\lambda}{\sqrt2} (\vl_2-\vk_1)
=\dfrac{e^\lambda}{\sqrt2} \left( \begin{array}{c} \hn_2 \\
-\hn_1  \end{array} \right), \nn\\
&D_\mu \vj=0,~~~D_\mu \vtj=0,
\label{b2ic}
\eea
To find the restricted connection $\hvGm$ which satisfies the
isometry condition we first introduce 4 more basis vectors in Lorentz
group manifold which together with $\vj$ and $\vtj$
form a complete basis
\bea
&\vk= \dfrac{e^{-\lambda}}{\sqrt 2} (\vl_1 -\vk_2)
=\dfrac{e^{-\lambda}}{\sqrt2} \left( \begin{array}{c} \hn_1 \\
-\hn_2 \end{array} \right),\nn\\
&\vtk= -\dfrac{e^{-\lambda}}{\sqrt 2} (\vl_2 +\vk_1)
=\dfrac{e^{-\lambda}}{\sqrt2} \left(\begin{array}{c}
-\hn_2 \\ -\hn_1 \end{array}\right), \nn\\
&\vl=-\vj \times \vtk=-\vtj \times \vk = \left( \begin{array}{c} \hn_3 \\
0 \end{array} \right),\nn\\
&\vtl=\vj \times \vk=-\vtj \times \vtk = \left( \begin{array}{c} 0 \\
-\hn_3 \end{array}\right).
\eea
Notice that 4 of them are null vectors,
\bea
&{\vj}^2= {\vtj}^2=\vk^2=\vtk^2=0,
\eea
but we have
\bea
&\vj \cdot \vk = -\vtj \cdot \vtk = 1,
~~~\vl^2=-\vtl^2=1.
\eea
All other scalar products of the basis vectors vanish.
Moreover we have
\bea
&\vj \times \vl= -\vtj \times \vtl =-\vtj,
~~~\vtj \times \vl=\vj \times \vtl=\vj, \nn\\
&\vk \times \vl= -\vtk \times \vtl =\vtk,
~~~\vtk \times \vl=\vk \times \vtl=-\vk.
\eea
From this we find the following restricted connection
for the $B_2$ isometry,
\bea
&\hvGm_\mu =\Gamma_\mu~\vj - \tGm_\mu~\vtj - \vk
\times \pro_\mu \vj  \nn\\
&= \Gamma_\mu~\vj - \tGm_\mu~\vtj
- \dfrac12(\vk \times \pro_\mu \vj-\vtk \times \pro_\mu \vtj), \nn\\
&\Gamma_\mu = \vk \cdot \vGm_\mu,
~~~\tGm_\mu = \vtk \cdot \vGm_\mu, \nn\\
&\vk \times \pro_\mu \vj=-\vtk \times \pro_\mu \vtj~,
\label{b2rc}
\eea
where $\Gamma_\mu$ and $\tGm_\mu$ are two Abelian connections
of $\vj$ and $\vtj$ components which are not restricted
by the isometry condition.

The restricted curvature tensor $\hvR_\mn$ is given by
\bea
&\hvR_\mn=\pro_\mu \hvGm_\nu-\pro_\nu \hvGm_\mu
+\hvGm_\mu \times \hvGm_\nu \nn\\
&=(\Gamma_\mn+H_\mn) \vj
-(\tGm_\mn+\tH_\mn) \vtj, \nn\\
& \Gamma_\mn = \pro_\mu \Gamma_\nu - \pro_\nu \Gamma_\mu,
~~~\tGm_\mn = \pro_\mu \tGm_\nu- \pro_\nu \tGm_\mu, \nn\\
& H_\mn = -\vk \cdot (\pro_\mu \vj \times \pro_\nu \vk
-\pro_\nu \vj \times \pro_\mu \vk), \nn\\
& \tH_\mn= -\vtk \cdot (\pro_\mu \vj \times \pro_\nu \vk -\pro_\nu
\vj \times \pro_\mu \vk),
\eea
so that
\bea
&\hR_\mn^{~~ab}=(\Gamma_\mn+H_\mn) j^{ab} -(\tGm_\mn+\tH_\mn)
\tj^{ab}.
\label{b2rct}
\eea
Notice that $\hvR_\mn$ is orthogonal to $\vl$ and $\vtl$.
This should be contrasted with the restricted curvature tensor
(\ref{a2rct}) of the $A_2$ isometry.

In 3-dimensional notation the isometry condition
(\ref{b2ic}) is written as
\bea
&\hvGm_\mu=\left( \begin{array}{c} \hA_\mu \\
\hB_\mu  \end{array} \right), \nn\\
&\hD_\mu \hn_1=\hB_\mu \times \hn_2-(\pro_\mu \lambda) \hn_1, \nn\\
&\hD_\mu \hn_2=-\hB_\mu \times \hn_1-(\pro_\mu \lambda) \hn_2.
\eea
From this we have
\bea
&\hA_\mu =A_\mu^1 \hn_1 + A_\mu^2 \hn_2
+(\hn_1 \cdot \pro_\mu \hn_2) \hn_3 \nn\\
&=\Big(\dfrac{e^{\lambda}}{\sqrt2} \Gamma_\mu
+\dfrac{\hn_2\cdot\pro_\mu\hn_3}{2}\Big) \hn_1
-\Big(\dfrac{e^{\lambda}}{\sqrt2} \tGm_\mu
-\dfrac{\hn_3\cdot\pro_\mu\hn_1}{2}\Big) \hn_2 \nn\\
&+(\hn_1 \cdot \pro_\mu \hn_2 )\hn_3,  \nn\\
&\hB_\mu = B_\mu^1 \hn_1 + B_\mu^2 \hn_2 - (\pro_\mu \lambda) \hn_3 \nn\\
&=\Big(\dfrac{e^{\lambda}}{\sqrt2} \tGm_\mu
+\dfrac{\hn_3\cdot\pro_\mu\hn_1}{2}\Big) \hn_1
+\Big(\dfrac{e^{\lambda}}{\sqrt2} \Gamma_\mu
-\dfrac{\hn_2\cdot\pro_\mu\hn_3}{2}\Big) \hn_2  \nn\\
&-(\pro_\mu \lambda)\hn_3,  \nn\\
&A_\mu^1=\dfrac{e^{\lambda}}{\sqrt2}\big(\Gamma_\mu-\tC_\mu^1 \big),
~~A_\mu^2=-\dfrac{e^{\lambda}}{\sqrt2}\big(\tGm_\mu-\tC_\mu^2 \big), \nn\\
&B_\mu^1=\dfrac{e^{\lambda}}{\sqrt2} \big(\tGm_\mu+\tC_\mu^2 \big),
~~B_\mu^2=\dfrac{e^{\lambda}}{\sqrt2}\big(\Gamma_\mu+\tC_\mu^1 \big),  \nn\\
&\tC_\mu^1=-\dfrac{e^{-\lambda}}{\sqrt 2}\hn_2\cdot\pro_\mu\hn_3, \nn\\
&\tC_\mu^2=-\dfrac{e^{-\lambda}}{\sqrt 2} \hn_1 \cdot\pro_\mu \hn_3,
\eea
so that
\bea
&\hA_\mu=-\hn_3 \times \hB_\mu
+\dfrac 12 \epsilon_{ijk}(\hn_i \cdot \pro_\mu \hn_j) \hn_k  \nn\\
&=B_\mu^2 \hn_1-B_\mu^1 \hn_2
+\dfrac 12 \epsilon_{ijk}(\hn_i \cdot \pro_\mu \hn_j) \hn_k,  \nn\\
&\hB_\mu=\hn_3 \times \hA_\mu-\pro_\mu \hn_3-(\pro_\mu \lambda) \hn_3 \nn\\
&=-A_\mu^2 \hn_1 +A_\mu^1 \hn_2-\pro_\mu \hn_3-(\pro_\mu \lambda) \hn_3.
\eea
Notice that both $\hA_\mu$ and $\hB_\mu$ have non-vanishing $\hn_3$
components.

With
\bea
\hvR_\mn=\left( \begin{array}{c} \hA_\mn \\
\hB_\mn  \end{array} \right) \nn
\eea
we have
\bea
&\hA_\mn=\pro_\mu \hA_\nu-\pro_\nu \hA_\mu
+\hA_\mu \times \hA_\nu-\hB_\mu \times \hB_\nu \nn\\
&=\dfrac{e^\lambda}{\sqrt 2}(\Gamma_\mn+H_\mn) \hn_1
-\dfrac{e^\lambda}{\sqrt 2}(\tGm_\mn+\tH_\mn) \hn_2 \nn\\
&=A_\mn^1 \hn_1 + A_\mn^2 \hn_2, \nn\\
&\hB_\mn=\pro_\mu \hB_\nu-\pro_\nu \hB_\mu
+\hA_\mu \times \hB_\nu-\hA_\nu \times \hB_\mu  \nn\\
&=\hD_\mu \hB_\nu-\hD_\nu \hB_\mu \nn\\
&=\dfrac{e^\lambda}{\sqrt 2}(\tGm_\mn+\tH_\mn) \hn_1
+\dfrac{e^\lambda}{\sqrt 2}(\Gamma_\mn+H_\mn) \hn_2  \nn\\
&=B_\mn^1 \hn_1 + B_\mn^2 \hn_2,
\eea
where
\bea
&H_\mn = \pro_\mu \tC_\nu^1 - \pro_\nu \tC_\mu^1
=\dfrac{e^{-\lambda}}{\sqrt2}
\Big(-\hn_1 \cdot(\pro_\mu \hn_1 \times \pro_\nu \hn_1) \nn\\
&+\hn_2 \cdot (\pro_\mu \lambda \pro_\nu \hn_3
-\pro_\nu \lambda \pro_\mu \hn_3) \Big), \nn\\
&\tH_\mn= \pro_\mu \tC_\nu^2 - \pro_\nu \tC_\mu^2
= \dfrac{e^{-\lambda}}{\sqrt2}
\Big(\hn_2 \cdot(\pro_\mu \hn_2 \times \pro_\nu \hn_2) \nn\\
&+\hn_1 \cdot (\pro_\mu \lambda \pro_\nu \hn_3 -\pro_\nu \lambda
\pro_\mu \hn_3) \Big), \nn\\
&A_\mn^1 = B_\mn^2 =\dfrac{e^\lambda}{\sqrt 2}
(\pro_\mu K_\nu - \pro_\nu K_\mu),  \nn\\
&A_\mn^2 = -B_\mn^1 =-\dfrac{e^\lambda}{\sqrt 2} (\pro_\mu
\tK_\nu - \pro_\nu \tK_\mu), \nn\\
&K_\mu=\Gamma_\mu +\tC_\mu^1,
~~~\tK_\mu=\tGm_\mu+\tC_\mu^2,
\eea
so that
\bea
&\hA_\mn=-\hn_3 \times \hB_\mn,
~~~\hB_\mn= \hn_3 \times \hA_\mn.
\eea
Notice that both $\hA_\mn$ and $\hB_\mn$ are orthogonal to $\hn_3$,
although $\hA_\mu$ and $\hB_\mu$ are not.

With this we obtain the full gauge potential of Lorentz group
by adding the valence connection $\vZ_\mu$,
\bea
& \vGm_\mu = \hvGm_\mu + \vZ_\mu, \nn\\
& \vk \cdot \vZ_\mu = \vtk \cdot \vZ_\mu = 0.
\label{b2vc}
\eea
With
\bea
&\vZ_\mu = J_\mu \vk -\tJ_\mu \vtk +L_\mu \vl -\tL_\mu \vtl,  \nn\\
&J_\mu=\vj \cdot \vZ_\mu,~~~\tJ_\mu=\vtj \cdot \vZ_\mu, \nn\\
&L_\mu=\vl \cdot \vZ_\mu,~~~\tL_\mu=\vtl \cdot \vZ_\mu,
\eea
we have
\bea
&\dZ_\mn= \hD_\mu \vZ_\nu-\hD_\nu \vZ_\mu \nn\\
&=U_\mn \vj -\tU_\mn \vtj +(\pro_\mu J_\nu-\pro_\nu J_\mu) \vk
-(\pro_\mu \tJ_\nu-\pro_\nu \tJ_\mu) \vtk  \nn\\
&+ (\cD_\mu L_\nu-\cD_\nu L_\mu) \vl
-(\cD_\mu \tL_\nu-\cD_\nu \tL_\mu) \vtl, \nn\\
&U_\mn = -K_\mu \tL_\nu -\tK_\mu L_\nu 
+(K_\nu \tL_\mu +\tK_\nu L_\mu), \nn\\
&\tU_\mn =K_\mu L_\nu -\tK_\mu \tL_\nu
-(K_\nu L_\mu -\tK_\nu \tL_\mu), \nn\\
&\cD_\mu L_\nu = \pro_\mu L_\nu +K_\mu \tJ_\nu+\tK_\mu J_\nu, \nn\\
&\cD_\mu \tL_\nu=\pro_\mu \tL_\nu-K_\mu J_\nu+\tK_\mu \tJ_\nu, \nn\\
&\pZ_\mn=\vZ_\mu \times \vZ_\nu=V_\mn \vk -\tV_\mn \vtk, \nn\\
&V_\mn=J_\mu \tL_\nu +\tJ_\mu L_\nu-(J_\nu \tL_\mu +\tJ_\nu L_\mu), \nn\\
&\tV_\mn=\tJ_\mu \tL_\nu-J_\mu L_\nu-(\tJ_\nu \tL_\mu-J_\nu L_\mu),
\eea
so that
\bea
&\vZ_\mn=\dZ_\mn+\pZ_\mn =U_\mn \vj -\tU_\mn \vtj \nn\\
&+(\cD_\mu J_\nu-\cD_\nu J_\mu)~\vk
-(\cD_\mu \tJ_\nu-\cD_\nu \tJ_\mu)~\vtk \nn\\
&+(\cD_\mu L_\nu-\cD_\nu L_\mu)~\vl
-(\cD_\mu \tL_\nu-\cD_\nu \tL_\mu)~\vtl, \nn\\
&\cD_\mu J_\nu = \pro_\mu J_\nu-\tL_\mu J_\nu-L_\mu \tJ_\nu, \nn\\
&\cD_\mu \tJ_\nu= \pro_\mu \tJ_\nu-\tL_\mu \tJ_\nu +L_\mu J_\nu.
\eea
Notice that in this case the kinetic part $\dZ_\mn$ contains
all six components, but the potential part $\pZ_\mn$ has only
$\vk$ and $\vtk$ components.
With this we have the full curvature tensor
\bea
&\vR_\mn =\hvR_\mn+\dZ_\mn+\pZ_\mn \nn\\
&=(\Gamma_\mn+H_\mn+U_\mn) \vj
-(\tGm_\mn+\tH_\mn+\tU_\mn) \vtj \nn\\
&+(\cD_\mu J_\nu-\cD_\nu J_\mu) \vk
-(\cD_\mu \tJ_\nu-\cD_\nu \tJ_\mu) \vtk \nn\\
&+(\cD_\mu L_\nu-\cD_\nu L_\mu) \vl
-(\cD_\mu \tL_\nu-\cD_\nu \tL_\mu) \vtl \nn\\
&=(\cD_\mu K_\nu-\cD_\nu K_\mu) \vj
-(\cD_\mu \tK_\nu-\cD_\nu \tK_\mu) \vtj \nn\\
&+(\cD_\mu J_\nu-\cD_\nu J_\mu) \vk
-(\cD_\mu \tJ_\nu-\cD_\nu \tJ_\mu) \vtk \nn\\
&+(\cD_\mu L_\nu-\cD_\nu L_\mu) \vl
-(\cD_\mu \tL_\nu-\cD_\nu \tL_\mu) \vtl \nn\\
&=K_\mn \vj - \tK_\mn \vtj + J_\mn \vk - \tJ_\mn \vtk 
+ L_\mn \vl - \tL_\mn \vtl, \nn\\
&\cD_\mu K_\nu= \pro_\mu K_\nu +\tL_\mu K_\nu +L_\mu \tK_\nu, \nn\\
&\cD_\mu \tK_\nu= \pro_\mu \tK_\nu +\tL_\mu \tK_\nu -L_\mu K_\nu, \nn\\
&K_\mn =\Gamma_\mn+H_\mn+U_\mn=\cD_\mu K_\nu-\cD_\nu K_\mu, \nn\\
&\tK_\mn =\tGm_\mn+\tH_\mn+\tU_\mn
=\cD_\mu \tK_\nu-\cD_\nu \tK_\mu, \nn\\
&J_\mn = \cD_\mu J_\nu-\cD_\nu J_\mu,~~\tJ_\mn= \cD_\mu \tJ_\nu-\cD_\nu \tJ_\mu, \nn\\
&L_\mn = \cD_\mu L_\nu-\cD_\nu L_\mu,~~\tL_\mn= \cD_\mu \tL_\nu-\cD_\nu \tL_\mu,
\label{b2ct1}
\eea
or equivalently
\bea
&R_\mn^{~~ab}=\vR_\mn \cdot \vI^{ab} \nn\\
&= K_\mn j^{ab} - \tK_\mn \tj^{ab} +J_\mn k^{ab} -\tJ_\mn \tk^{ab} \nn\\
&+ L_\mn l^{ab} - \tL_\mn \tl^{ab},
\label{b2ct2}
\eea
This is the $B_2$ decomposition of the curvature tensor.

With complex notation
\bea
&\vk_\pm=\dfrac{1}{\sqrt2}(\vk\pm i\vl),
~~~~\vtk_\pm=\dfrac{1}{\sqrt2}(\vtk\pm i\vtl), \nn\\
&Z_\mu=\dfrac{1}{\sqrt2}(J_\mu+iL_\mu),
~~~~\tZ_\mu= \dfrac{1}{\sqrt2}(\tJ_\mu+i\tL_\mu), \nn\\
&Z_\mu'=\dfrac{1}{\sqrt2}(K_\mu+iL_\mu)
=Z_\mu-\dfrac{1}{\sqrt2} B^-_\mu, \nn\\
&\tZ_\mu'= \dfrac{1}{\sqrt2}(\tK_\mu+i\tL_\mu)
=\tZ_\mu-\dfrac{1}{\sqrt2}\tB^-_\mu,  \nn\\
&B^\pm_\mu=J_\mu \pm K_\mu,~~~\tB^\pm_\mu=\tJ_\mu \pm \tK_\mu,
\eea
we obtain
\bea
&\vZ_\mn= i(\tZ'^*_\mu Z'_\nu-\tZ'^*_\nu Z'_\mu
+Z'^*_\mu \tZ'_\nu- Z'^*_\nu \tZ'_\mu)\vj \nn\\
&+i(\tZ'_\mu \tZ'^*_\nu - \tZ'_\nu \tZ'^*_\mu 
+Z'^*_\mu Z'_\nu - Z'^*_\nu Z'_\mu)\vtj \nn\\
&+(\cD_\mu Z_\nu-\cD_\nu Z_\mu)^*\vk_+
-(\cD_\mu \tZ_\nu-\cD_\nu \tZ_\mu)^*\vtk_+, \nn\\
&+(\cD_\mu Z_\nu-\cD_\nu Z_\mu)\vk_-
-(\cD_\mu \tZ_\nu-\cD_\nu \tZ_\mu)\vtk_-, \nn\\
&\cD_\mu Z_\nu= \pro_\mu Z_\nu -\dfrac{i}{2}(\tB^-_\mu Z_\nu -\tB^+_\mu Z_\nu^* \nn\\
&+B^-_\mu \tZ_\nu -B^+_\mu \tZ_\nu^*),  \nn\\
&\cD_\mu \tZ_\nu=\pro_\mu \tZ_\nu -\dfrac{i}{2}(\tB^-_\mu \tZ_\nu -\tB^+_\mu \tZ_\nu^* \nn\\
&-B^-_\mu Z_\nu +B^+_\mu Z_\nu^* ).
\eea
With this we have
\bea
&\vR_\mn = (\cD_\mu K_\nu-\cD_\nu K_\mu)~\vj
-(\cD_\mu \tK_\nu-\cD_\nu \tK_\mu)~\vtj \nn\\
&+(\cD_\mu Z_\nu-\cD_\nu Z_\mu)^*~\vk_+
-(\cD_\mu \tZ_\nu-\cD_\nu \tZ_\mu)^*~\vtk_+  \nn\\
&+(\cD_\mu Z_\nu-\cD_\nu Z_\mu)~\vk_- \nn\\
&-(\cD_\mu \tZ_\nu-\cD_\nu \tZ_\mu)~\vtk_-,
\label{b2ct3}
\eea
or
\bea
&R_\mn^{~~ab}=(\cD_\mu K_\nu-\cD_\nu K_\mu)~j^{ab}
-(\cD_\mu \tK_\nu-\cD_\nu \tK_\mu)~\tj^{ab}  \nn\\
&+(\cD_\mu Z_\nu-\cD_\nu Z_\mu)^*~k_+^{ab}
-(\cD_\mu \tZ_\nu-\cD_\nu \tZ_\mu)^*~\tk_+^{ab} \nn\\
&+(\cD_\mu Z_\nu-\cD_\nu Z_\mu)~k_-^{ab} \nn\\
&-(\cD_\mu \tZ_\nu-\cD_\nu \tZ_\mu)~\tk_-^{ab}.
\label{b2ct4}
\eea
This should be compared with the $A_2$ result (\ref{a2ct3}) or
(\ref{a2ct4}).

In 3-dimensional notation, we have
\bea
&\vZ_\mu=\left( \begin{array}{c} \X_\mu \\
\Y_\mu  \end{array} \right), \nn\\
&\X_\mu =\dfrac{e^{-\lambda}}{\sqrt2} \big( J_\mu \hn_1
+\tJ_\mu \hn_2 \big)+L_\mu \hn_3,  \nn\\
&\Y_\mu=\dfrac{e^{-\lambda}}{\sqrt2} \big( \tJ_\mu \hn_1
-J_\mu \hn_2 \big)+\tL_\mu \hn_3,
\eea
so that
\bea
&\hn_1 \cdot \X_\mu +\hn_2 \cdot \Y_\mu=0,  \nn\\
&\hn_2 \cdot \X_\mu -\hn_1 \cdot \Y_\mu=0,  \nn\\
&\hn_3 \times \Y_\mu= -\hn_3 \times (\hn_3 \times \X_\mu).
\eea
Moreover, with
\bea
&\vZ_\mn=\left( \begin{array}{c} \X_\mn \\
\Y_\mn  \end{array} \right)
=\left( \begin{array}{c} \dX_\mn+\pX_\mn \\
\dY_\mn+\pY_\mn  \end{array} \right),
\eea
we have
\bea
&\dX_\mn=\Big\{\dfrac{e^{\lambda}}{\sqrt2} U_\mn
+\dfrac{e^{-\lambda}}{\sqrt2} (\pro_\mu J_\nu
-\pro_\nu J_\mu) \Big\} \hn_1 \nn\\
&-\Big\{\dfrac{e^{\lambda}}{\sqrt2}\tU_\mn
-\dfrac{e^{-\lambda}}{\sqrt2} (\pro_\mu \tJ_\nu
-\pro_\nu \tJ_\mu) \Big\} \hn_2 +L_\mn \hn_3, \nn\\
&\dY_\mn=\Big\{\dfrac{e^{\lambda}}{\sqrt2}\tU_\mn
+\dfrac{e^{-\lambda}}{\sqrt2} (\pro_\mu \tJ_\nu
-\pro_\nu \tJ_\mu) \Big\}\hn_1 \nn\\
&+\Big\{\dfrac{e^{\lambda}}{\sqrt2} U_\mn
-\dfrac{e^{-\lambda}}{\sqrt2} (\pro_\mu J_\nu
-\pro_\nu J_\mu) \Big\}\hn_2 +\tL_\mn \hn_3, \nn\\
&\pX_\mn=\dfrac{e^{-\lambda}}{\sqrt2} ( V_\mn \hn_1 +\tV_\mn \hn_2), \nn\\
&\pY_\mn=\dfrac{e^{-\lambda}}{\sqrt2} (\tV_\mn \hn_1-V_\mn \hn_2),
\eea
so that
\bea
&\X_\mn=\Big(\dfrac{e^{\lambda}}{\sqrt2}U_\mn
+\dfrac{e^{-\lambda}}{\sqrt2}J_\mn \Big)\hn_1 \nn\\
&-\Big(\dfrac{e^{\lambda}}{\sqrt2}\tU_\mn
-\dfrac{e^{-\lambda}}{\sqrt2} \tJ_\mn \Big)\hn_2 +L_\mn \hn_3, \nn\\
&\Y_\mn=\Big(\dfrac{e^{\lambda}}{\sqrt2}\tU_\mn
+\dfrac{e^{-\lambda}}{\sqrt2} \tJ_\mn \Big)\hn_1 \nn\\
&+\Big(\dfrac{e^{\lambda}}{\sqrt2} U_\mn
-\dfrac{e^{-\lambda}}{\sqrt2} J_\mn \Big)\hn_2 +\tL_\mn \hn_3.
\eea
Finally with
\bea
&\vR_\mn =\left(\begin{array}{c} \A_\mn \\
\B_\mn  \end{array} \right)=\left(\begin{array}{c} \hA_\mn+\X_\mn \\
\hB_\mn+\Y_\mn  \end{array} \right),
\eea
we have
\bea
&\A_\mn=\Big(\dfrac{e^{\lambda}}{\sqrt2}K_\mn
+\dfrac{e^{-\lambda}}{\sqrt2}J_\mn \Big)\hn_1 \nn\\
&-\Big(\dfrac{e^{\lambda}}{\sqrt2}\tK_\mn
-\dfrac{e^{-\lambda}}{\sqrt2} \tJ_\mn \Big)\hn_2 +L_\mn \hn_3, \nn\\
&\B_\mn=\Big(\dfrac{e^{\lambda}}{\sqrt2}\tK_\mn
+\dfrac{e^{-\lambda}}{\sqrt2} \tJ_\mn \Big)\hn_1 \nn\\
&+\Big(\dfrac{e^{\lambda}}{\sqrt2} K_\mn
-\dfrac{e^{-\lambda}}{\sqrt2} J_\mn \Big)\hn_2
+\tL_\mn \hn_3.
\eea
This completes the $B_2$ decomposition of the gravitational
connection.

The above result tells that there exist two different
Abelian decompositions of the gravitational connection
and the curvature tensor which decompose them
into the restricted part and the valence part.
This allows us to decompose the Einstein's
theory in terms of the restricted part
and the valence part.

\section{Abelian Decomposition of Einstein's Theory}

Now we are ready to discuss the decomposition of
Einstein's theory. Since the Einstein-Hilbert action is
described by the metric we have to express the above
decomposition of the gravitational
connection in terms of the metric. To do this we use the first order
formalism of Einstein theory. In the absence of the matter field,
the Einstein-Hilbert action in the first order
formalism is given by
\bea
&S[e^\mu_a,~\vGm_\mu]=\dfrac{1}{16\pi G_N} \int e~\Big(e^\mu_a~e^\nu_b
~\vI^{ab} \cdot \vR_\mn \Big)~d^4x  \nn\\
&=\dfrac{1}{16\pi G_N} \int e \Big(\vg_\mn \cdot \vR^\mn\Big)~d^4x, \nn\\
&e= {\rm Det}~(e_{\mu a}),~~~~~\vg_\mn = e_\mu^a~e_\nu^b~\vI_{ab}, \nn\\
&g_\mn^{~~ab}=(e_\mu^{~a} e_\nu^{~b}
-e_\nu^{~a} e_\mu^{~b})=g_{[\mn]}^{~~~~[ab]}. 
\label{Elag}
\eea
Here we have introduced the Lorentz covariant
four index metric tensor $\vg_\mn$ (which should not be confused with
the two index space-time metric $g_\mn$) which forms
an adjoint representation of Lorentz group. Notice that $\vg_\mn$ is 
antisymmetric in $\mu$ and $\nu$. Clearly this Lorentz covariant
metric becomes the natural metric which plays the role of $g_\mn$ in this 
gauge formalism. 

From (\ref{Elag}) we have the following equation of motion
\bea
&\delta e_{\mu a};~~~\vg_\mn \cdot \vR^{\nu\rho} e_{\rho a}
=R_{\mu a}=0 \nn\\
&\delta \vGm_\mu;~~~{\mathscr D}_\mu \vg^\mn
= (\nabla_\mu +\vGm_\mu \times) \vg^\mn=0,
\label{Eeq}
\eea
where $R_{\mu a}=e^{\nu b} R_{\mn ab}$ is the Ricci tensor and 
${\mathscr D}_\mu$ is generally and gauge covariant derivative.
Clearly the first equation is nothing but the Einstein's
equation in the absence of matter field. But
the Ricci tensor is written in terms of the gauge potential,
not the metric. 

To understand the meaning of the second equation,
notice that the second equation tells that $\vg_\mn$
is invariant under the parallel transport along the
$\pro_\mu$-direction defined by the gauge potential $\vGm_\mu$,
which puts a strong constraint on $\vGm_\mu$.
In fact from this one can show that $\vGm_\mu$ is given by
\bea
&\vGm_\mu \cdot \vI^{ab}=\dfrac12 (e^{a\nu} e_{c\mu} \pro^b e^c_{~\nu}
+e^{a\nu}\pro_\mu e^b_{~\nu}+\pro^b e^a_{~\mu} \nn\\
&-e^{b\nu} e_{c\mu} \pro^a e^c_{~\nu}
-e^{b\nu}\pro_\mu e^a_{~\nu}-\pro^a e^b_{~\mu})=\Gm_\mu^{~ab}.
\label{scon}
\eea
But this, of course, is the Levi-Civita connection written in the tetrad 
basis. This confirms that the gauge potential $\vGm_\mu$ of Lorentz group 
becomes the (torsion-free) spin connection $\omega_\mu^{~ab}$, which assures 
that (\ref{Eeq}) indeed describes the Einstein's general relativity.
But remember that in general it can have torsion 
when a spinor source is present \cite{prd76b}.

This telle that the second equation of (\ref{Eeq}) is nothing but 
the metric-compatibility condition of the connection
\bea
&{\mathscr D}_\mu \vg^\mn=0 \Longleftrightarrow \nabla_\alpha g_\mn=0. 
\eea
But actually in the Lorentz gauge formalizm of Einstein's theory we 
have this metric-compatibility from the beginning, because we already have 
\bea
D_\mu \eta_{ab}=0.
\label{invm}
\eea 
Indeed, with this and with the identity
\bea
{\mathscr D}_\mu e_\nu^{~a}=\pro_\mu e_\nu^{~a}-\Gm_\mn^{~~\alpha} e_\alpha^{~a}
+\Gm_{\mu~b}^{~a} e_\nu^{~b}=0,
\eea
$\mathscr D_\mu \vg^\mn$ is reduced to
\bea
D_\mu \vI^{ab}=0,
\eea
which becomes an identity with (\ref{invm}). So the second equation of (\ref{Eeq})
can actually be viewed as an identity.

\subsection{$A_2$ (Non Light-like) Decomposition}

With this preliminary, we discuss the decomposition of
Einstein's theory with the $A_2$ isometry
(the space/time isometry) first.
For this we introduce two projection operators which project out
the isometry components,
\bea
&\vSi_{ab}=l_{ab}~\vl-\tl_{ab}~\vtl, \nn\\
& \vP_{ab}=\vI_{ab}-\vSi_{ab}=l^1_{ab}~\vl_1-\tl^1_{ab}~\vtl_1
+l^2_{ab}~\vl_2-\tl^2_{ab}~\vtl_2, \nn\\
&\Si_{ab}^{~~cd}=l_{ab} l^{cd}-\tl_{ab} \tl^{cd},
~~~~\Pi_{ab}^{~~cd}=I_{ab}^{~~cd}-\Si_{ab}^{~~cd},  \nn\\
&\vZ_\mu \cdot \vSi_{ab}=0,~~~\vZ_\mu \cdot \vP_{ab}=Z_\mu^{~ab}.
\eea
Clearly $\vSi_{ab}$ and $\vP_{ab}$ become projection operators
in the sense that
\bea
&\vSi_{ab} \cdot \vSi^{cd} =\dfrac 12 \Si_{ab}^{~~mn}\Si_{mn}^{~~cd}
=\Si_{ab}^{~~cd}, \nn\\
&\vP_{ab} \cdot \vP^{cd}=\dfrac 12 \Pi_{ab}^{~~mn}\Pi_{mn}^{~~cd}
=\Pi_{ab}^{~~cd},  \nn\\
&\vSi_{ab} \cdot \vP^{cd}=0.
\eea
Now we can express the Einstein-Hilbert action as
\bea
&S[e^\mu_a,~A_\mu,~B_\mu,~\vZ_\mu]
=\dfrac{1}{16\pi G_N} \int e~\Big\{~\vg_\mn \cdot \vR^\mn \nn \\
&+\lambda (\vl^2-1) + \tilde \lambda (\vl \cdot \vtl)
+\lambda_\mu (\vl \cdot \vZ^\mu)
+\tilde \lambda_\mu (\vtl \cdot \vZ^\mu)\Big\}~d^4x, \nn\\
&\vR_\mn =\hvR_\mn+(\hD_\mu \vZ_\nu-\hD_\nu \vZ_\mu)
+\vZ_\mu \times \vZ_\nu  \nn\\
&=(\cD_\mu \bA_\nu-\cD_\nu \bA_\mu) \vl
-(\cD_\mu B_\nu-\cD_\nu B_\mu) \vtl \nn\\
&+ (\hD_\mu \vZ_\nu-\hD_\nu \vZ_\mu),
\label{a2lag1}
\eea
where $\lambda's$ are the Lagrange multipliers. From this we
get the following equations of motion
\bea
&\delta e_{\mu c};~~(e_\mu^a~e_\nu^b)
\big[(\cD^\nu \bA^\rho-\cD^\rho \bA^\nu)~l_{ab} \nn\\
&-(\cD^\nu B^\rho-\cD^\rho B^\nu)\tl_{ab}
+(\hD^\nu \vZ^\rho-\hD^\rho \vZ^\nu) \cdot \vP_{ab} \big] e_{\rho c} \nn\\
&=0, \nn\\
&\delta A_\nu;~~ \nabla_\mu (e_a^\mu ~e_b^\nu ~l^{ab})
+\vl  \cdot ( \vZ_\mu \times \vg^\mn)=0,  \nn\\
& \delta B_\nu;~~\nabla_\mu (e_a^\mu~e_b^\nu~\tl^{ab})
+\vtl  \cdot ( \vZ_\mu \times \vg^\mn)=0,  \nn\\
& \delta \vZ_\nu;~~\hat{{\mathscr D}}_\mu (e_a^\mu ~e_b^\nu ~\vP^{ab})
+(e_a^\mu e_b^\nu) \big[(\vZ_\mu \times \vl)l^{ab}  \nn\\ 
&-(\vZ_\mu \times \vtl) \tl^{ab} \big]=0.  \nn\\
&\hat{{\mathscr D}}_\mu=\nabla_\mu+\hvGm_\mu \times.
\label{a2Eeq1}
\eea
Notice that, using the isometry (\ref{a2ic}), we can
combine the last three equations into a single equation,
\bea
{\mathscr D}_\mu \vg^\mn=0.
\eea
But this is precisely the second equation of (\ref{Eeq}),
which confirms that (\ref{a2Eeq1}) is equivalent to (\ref{Eeq}).

To clarify the meaning of the above equation
we define the restricted metric $\hvg_\mn$ decomposing
$\vg_\mn$
\bea
&\vg_\mn = \hvg_\mn + \vG_\mn,  \nn\\
&\hvg_\mn =e_\mu^a~e_\nu^b~\vSi_{ab}=G_\mn ~\vl -\tG_\mn \vtl, \nn\\
&\vG_\mn= e_\mu^a~e_\nu^b~\vP_{ab}=G_\mn^1 \vl_1-\tG_\mn^1 \vtl_1 
+G_\mn^2 \vl_2-\tG_\mn^2 \vtl_2, \nn\\
&G_\mn=e_\mu^a~e_\nu^b~l_{ab},
~~~\tG_\mn = e_\mu^a~e_\nu^b~\tl_{ab}, \nn\\
&G_\mn^1=e_\mu^a~e_\nu^b~l_{ab}^1,
~~~\tG_\mn^1 = e_\mu^a~e_\nu^b~\tl_{ab}^1, \nn\\
&G_\mn^2=e_\mu^a~e_\nu^b~l_{ab}^2,
~~~\tG_\mn^2 = e_\mu^a~e_\nu^b~\tl_{ab}^2.
\label{gdef}
\eea
Notice that
\bea
&\tG_\mn = \dfrac 12 \epsilon_{abcd} e_\mu^{~a} ~e_\nu^{~b}
~l^{cd}
=\dfrac 12 \epsilon_{\mn cd} l^{cd} \nn\\
&= \dfrac 12 \epsilon_{\mn\rho\sigma} G^{\rho\sigma} = G_\mn^d,  \nn\\
&\tG_\mn^1 = G_\mn^{1~d},~~~~~\tG_\mn^2 = G_\mn^{2~d}.
\eea
Clearly the two two-forms $G_\mn$ and $\tG_\mn$ can be viewed
to represent the restricted metric which are dual to each other.
With this (\ref{a2Eeq1}) has the following compact expression
\bea
&G_\mn (\cD^\nu \bA^\rho-\cD^\rho \bA^\nu)
-\tG_\mn (\cD^\nu B^\rho-\cD^\rho B^\nu) \nn\\
&+\vG_\mn \cdot (\hD^\nu \vZ^\rho-\hD^\rho \vZ^\nu)=0, \nn\\
& \nabla_\mu G^\mn +\vl \cdot (\vZ_\mu \times \vG^\mn) =0, \nn\\
& \nabla_\mu \tG^\mn +\vtl \cdot (\vZ_\mu \times \vG^\mn) =0, \nn\\
& \hat{{\mathscr D}}_\mu \vG^\mn +\vZ_\mu \times \hvg^\mn=0,
\label{a2Eeq2}
\eea
or equivalently
\bea
&G_\mn (\cD^\nu \bA^\rho-\cD^\rho \bA^\nu)
-\tG_\mn (\cD^\nu B^\rho-\cD^\rho B^\nu) \nn\\
&+G_\mn^i (\cD^\nu Z^\rho_i -\cD^\rho Z^\nu_i)
-\tG_\mn^i (\cD^\nu \tZ^\rho_i-\cD^\rho \tZ^\nu_i)=0, \nn\\
&\nabla_\mu G^\mn+\epsilon_{ij} (Z^i_\mu G_j^\mn-\tZ^i_\mu \tG_j^\mn)=0, \nn\\
&\nabla_\mu \tG^\mn+\epsilon_{ij} (Z^i_\mu \tG_j^\mn+\tZ^i_\mu G_j^\mn)=0, \nn\\
& \nabla_\mu G_i^\mn -\epsilon_{ij}(\bA_\mu G_j^\mn -B_\mu
\tG_j^\mn - Z^j_\mu G^\mn +\tZ^j_\mu \tG^\mn)  \nn\\
&=0, \nn\\
& \nabla_\mu \tG_i^\mn -\epsilon_{ij}(\bA_\mu \tG_j^\mn +B_\mu
G_j^\mn -Z^j_\mu \tG^\mn - \tZ^j_\mu G^\mn)  \nn\\
&=0. \nn\\
&(i,j=1,2,~~~\epsilon_{12}=-\epsilon_{21}=1)
\label{a2Eeq3}
\eea
This suggests that the valence connection $\vZ_\mu$
plays the role of the gravitational source of
the restricted metric.

In 3-dimensional notation we have
\bea
&\hvg_\mn=\left( \begin{array}{c} \hm_\mn \\
\he_\mn  \end{array} \right),
~~~\vG_\mn=\left( \begin{array}{c} \M_\mn \\
\E_\mn  \end{array} \right), \nn\\
&\vg_\mn=\left( \begin{array}{c} \hm_\mn+\M_\mn \\
\he_\mn+\E_\mn  \end{array} \right), \nn\\
&\hm_\mn = G_\mn \hn,~~~\he_\mn = \tG_\mn \hn, \nn\\
&\M_\mn = G^1_\mn \hn_1+G^2_\mn \hn_2, \nn\\ 
&\E_\mn = \tG^1_\mn \hn_1+\tG^2_\mn \hn_2,
\eea
so that the Einstein-Hilbert action (\ref{a2lag1}) acquires
the following form
\bea
&S[e^\mu_a,~A_\mu,~B_\mu,~Z^i_\mu,~\tZ^i_\mu] \nn\\
&= \dfrac{1}{16\pi G_N} \int e~\Big\{G_\mn (\cD^\mu \bA^\nu-\cD^\nu \bA^\mu) \nn\\
&-\tG_\mn (\cD^\mu B^\nu-\cD^\nu B^\mu)
+G_\mn^i (\cD^\mu Z^\nu_i -\cD^\nu Z^{\mu}_i)  \nn\\
&-\tG_\mn^i (\cD^\mu \tZ^{\nu}_i-\cD^\nu \tZ^{\mu}_i) \Big\} d^4x.
\label{a2lag2}
\eea
From this we can reproduce (\ref{a2Eeq3}).
This completes the $A_2$ decomposition (the space-like
decomposition) of Einstein's theory.

\subsection{$B_2$ (Light-like) Decomposition}

We can repeat the same procedure with the $B_2$ isometry
(the null isometry) to obtain
the desired decomposition of Einstein's equation.
With the Einstein-Hilbert action
\bea
&S[e^\mu_a,~\Gm_\mu,~\tGm_\mu,~\vZ_\mu]
=\dfrac{1}{16\pi G_N} \int
e~\Big\{\vg_\mn \cdot~\vR_\mn  \nn \\
&+\lambda~\vj^2 +\tilde \lambda (\vj \cdot \vtj)
+\lambda_\mu (\vk \cdot \vZ^\mu)
+\tilde \lambda_\mu (\vtk \cdot \vZ^\mu)\Big\}~d^4x,\nn\\
&\vR_\mn =\hvR_\mn+(\hD_\mu \vZ_\nu-\hD_\nu \vZ_\mu)
+\vZ_\mu \times \vZ_\nu  \nn\\
&=(\cD_\mu K_\nu-\cD_\nu K_\mu)~\vj
-(\cD_\mu \tK_\nu-\cD_\nu \tK_\mu)~\vtj  \nn\\
&+(\cD_\mu J_\nu-\cD_\nu J_\mu) \vk
-(\cD_\mu \tJ_\nu-\cD_\nu \tJ_\mu) \vtk \nn\\
&+(\cD_\mu L_\nu-\cD_\nu L_\mu) \vl
-(\cD_\mu \tL_\nu-\cD_\nu \tL_\mu) \vtl,
\label{b2lag1}
\eea
we get following equations of motion
\bea
& \delta e_{\mu c} ~;~(e^a_\mu ~e^b_\nu)
\Big[ (\cD^\nu K^\rho-\cD^\rho K^\nu)~j_{ab} \nn\\
&-(\cD^\nu \tK^\rho-\cD^\rho \tK^\nu)~\tj_{ab}
+\vZ^{\nu\rho} \cdot \vP_{ab} \Big] e_{\rho c}=0, \nn\\
&\delta \Gamma_\nu ~;~ \nabla_\mu (e_a^\mu ~e_b^\nu ~j^{ab})
+\vj \cdot (\vZ_\mu \times \vg^\mn)=0, \nn\\
&\delta \tGm_\nu ~;~ \nabla_\mu (e_a^\mu ~e_b^\nu ~\tj^{ab})
+\vtj \cdot (\vZ_\mu \times \vg^\mn)=0, \nn\\
&\delta \vZ_\nu ~;~ \hat{{\mathscr D}}_\mu (e_a^\mu ~e_b^\nu ~\vP^{ab})
+ \vZ_\mu \times (e_a^\mu ~e_b^\nu)(k^{ab}\vj -\tk^{ab} \vtj ) \nn\\
&=(e_a^\mu ~e_b^\nu) (j^{ab} \hD_\mu \vk-\tj^{ab} \hD_\mu \vtk),
\label{b2Eeq1}
\eea
where now
\bea
&\vSi_{ab}=j_{ab}~\vk-\tj_{ab}~\vtk,  \nn\\
&\vP_{ab}=k_{ab}~\vj-\tk_{ab}~\vtj +l_{ab}~\vl-\tl_{ab}~\vtl
=\vI_{ab}-\vSi_{ab},  \nn\\
&\vZ_\mu \cdot \vSi^{ab}=0,~~~\vZ_\mu \cdot \vP^{ab}=Z_\mu^{~ab}.
\eea
But notice that here $\vP_{ab}$ and $\vSi_{ab}$ do not make
projection operators, because
\bea
&\vP_{ab} \cdot \vSi^{cd}=k_{ab}~j^{cd}-\tk_{ab}~\tj^{cd} \neq 0.
\eea
Now, again we can combine the last three equations of
(\ref{b2Eeq1}) into a single equation with the isometry (\ref{b2ic}),
\bea
{\mathscr D}_\mu \vg^\mn=0. \nn
\eea
This confirms that (\ref{b2Eeq1}) is equivalent to (\ref{Eeq}),
which tells that (\ref{b2lag1}) describes the Einstein's gravity.

Now, with
\bea
&\vg_\mn = \hvg_\mn +\vG_\mn, \nn\\
&\hvg_\mn=e_\mu^a~e_\nu^b~\vSi^{ab}=\cJ_\mn~\vk-\ctJ_\mn~\vtk, \nn\\
&\vG_\mn= e_\mu^a~e_\nu^b~\vP^{ab} 
=\cK_\mn~\vj-\ctK_\mn~\vtj +\cL_\mn~\vl-\ctL_\mn~\vtl,  \nn\\
&\cJ_\mn=e_\mu^a~e_\nu^b~j_{ab},
~~~\ctJ_\mn=e_\mu^a~e_\nu^b~\tj_{ab}, \nn\\
&\cK_\mn=e_\mu^a~e_\nu^b~k_{ab},
~~~\ctK_\mn=e_\mu^a~e_\nu^b~\tk_{ab}, \nn\\
&\cL_\mn=e_\mu^a~e_\nu^b~l_{ab},
~~~\ctL_\mn=e_\mu^a~e_\nu^b~\tl_{ab},
\eea
the equation (\ref{b2Eeq1}) is written as
\bea
&\cJ_\mn (\cD^\nu K^\rho-\cD^\rho K^\nu)
-\ctJ_\mn (\cD^\nu \tK^\rho-\cD^\rho \tK^\nu) \nn\\
&+\vG_\mn \cdot \vZ^{\nu\rho} =0, \nn\\
& \nabla_\mu \cJ^\mn +\vj \cdot (\vZ_\mu \times \vg^\mn)=0, \nn\\
& \nabla_\mu \ctJ^\mn +\vtj \cdot (\vZ_\mu \times \vg^\mn)=0, \nn\\
& \hat{{\mathscr D}}_\mu \vG^\mn +\vZ_\mu \times (\cK^\mn~\vj-\ctK^\mn~\vtj) \nn\\
&=-\cJ^\mn \hD_\mu \vk+\ctJ^\mn \hD_\mu \vtk,
\label{b2Eeq2}
\eea
or equivalently
\bea
&\cJ_\mn (\cD^\nu K^\rho-\cD^\rho K^\nu)
-\ctJ_\mn (\cD^\nu \tK^\rho-\cD^\rho \tK^\nu) \nn\\
&+\cK_\mn (\cD^\nu J^\rho-\cD^\rho J^\nu)
-\ctK_\mn (\cD^\nu \tJ^\rho-\cD^\rho \tJ^\nu) \nn\\
&+\cL_\mn (\cD^\nu L^\rho-\cD^\rho L^\nu)
-\ctL_\mn (\cD^\nu \tL^\rho-\cD^\rho \tL^\nu)=0, \nn\\
& \nabla_\mu \cJ^\mn -L_\mu \ctJ^\mn -\tL_\mu \cJ^\mn
+J_\mu \ctL^\mn +\tJ_\mu \cL^\mn=0, \nn\\
& \nabla_\mu \ctJ^\mn +L_\mu \cJ^\mn -\tL_\mu \ctJ^\mn
-J_\mu \cL^\mn +\tJ_\mu \ctL^\mn=0, \nn\\
& \nabla_\mu \cK^\mn +L_\mu \ctK^\mn +\tL_\mu \cK^\mn 
=K_\mu \ctL^\mn +\tK_\mu \cL^\mn, \nn\\
& \nabla_\mu \ctK^\mn -L_\mu \cK^\mn +\tL_\mu \ctK^\mn 
=-K_\mu \cL^\mn +\tK_\mu \ctL^\mn, \nn\\
& \nabla_\mu \cL^\mn -J_\mu \ctK^\mn -\tJ_\mu \cK^\mn 
=-K_\mu \ctJ^\mn -\tK_\mu \cJ^\mn, \nn\\
& \nabla_\mu \ctL^\mn +J_\mu \cK^\mn -\tJ_\mu \ctK^\mn \nn\\
&= K_\mu \cJ^\mn-\tK_\mu \ctJ^\mn.
\label{b2Eeq3}
\eea
Remember that $\cJ_\mn,~\cK_\mn,~\cL_\mn$ and
$\ctJ_\mn,~\ctK_\mn,~\ctL_\mn$ are dual
to each other. Here again the valence connection becomes
the gravitational source of the restricted metric.

In 3-dimensional notation we have
\bea
&\hvg_\mn=\left( \begin{array}{c} \hm_\mn \\
\he_\mn  \end{array} \right),
~~~\vG_\mn=\left( \begin{array}{c} \M_\mn \\
\E_\mn  \end{array} \right), \nn\\
&\vg_\mn=\left( \begin{array}{c} \hm_\mn+\M_\mn \\
\he_\mn+\E_\mn  \end{array} \right), \nn\\
&\hm_\mn = \dfrac{e^{-\lambda}}{\sqrt2} (\cJ_\mn \hn_1+\ctJ_\mn \hn_2)
= \hn_3 \times \he_\mn, \nn\\
&\he_\mn = \dfrac{e^{-\lambda}}{\sqrt2} (\ctJ_\mn \hn_1-\cJ_\mn \hn_2)
=-\hn_3 \times \hm_\mn,  \nn\\
&\M_\mn = \dfrac{e^{\lambda}}{\sqrt2} (\cK_\mn \hn_1-\ctK_\mn \hn_2)
+\cL_\mn \hn_3, \nn\\
&\E_\mn = \dfrac{e^{\lambda}}{\sqrt2} (\ctK_\mn \hn_1+ \cK_\mn
\hn_2)+\ctL_\mn \hn_3,
\eea
so that the Einstein-Hilbert action (\ref{b2lag1})
is expressed as
\bea
&S[e^\mu_a,~K_\mu,~\tK_\mu,~J_\mu,~\tJ_\mu,~L_\mu,~\tL_\mu] \nn\\
&= \dfrac{1}{16\pi G_N} \int
e~\Big\{ \cJ_\mn (\cD^\mu K^\nu-\cD^\nu K^\mu)  \nn\\
&-\ctJ_\mn (\cD^\mu \tK^\nu-\cD^\nu \tK^\mu)
+\cK_\mn (\cD^\mu J^\nu-\cD^\nu J^\mu) \nn\\
&-\ctK_\mn(\cD^\mu \tJ^\nu-\cD^\nu \tJ^\mu)
+\cL_\mn (\cD^\mu L^\nu-\cD^\nu L^\mu) \nn\\
&-\ctL_\mn (\cD^\mu \tL^\nu-\cD^\nu \tL^\mu) \Big\}~d^4x.
\label{b2lag2}
\eea
From this we can reproduce (\ref{b2Eeq3}). 
This completes the $B_2$ decomposition (the light-like decomposition) of 
Einstein's theory.

\section{Restricted Gravity}

So far our analysis has been mainly on mathematical formalism, and one might 
wonder what is the physics behind it. The physical motivation behind the 
mathematical formalism is that we can simplify the Einstein's gravitation 
and obtain a restricted gravity which can describe the core dynamics of 
Einstein's theory without compromising the general invariance. In particular 
the Abelian decomposition allows us to describe the dynamical degrees of 
Einstein's theory by a spin-one Abelian gauge field.  

A common difficulty in quantum gravity and in non-Abelian quantum gauge theory
is the highly non-linear self interaction. In gauge theory one can simplify this 
non-linear interaction by separating the gauge covariant valence part from the
Abelian part of the potential and making the Abelian projection to obtain 
the restricted gauge theory \cite{prd80,prl81}. Here we can simplify Einstein's 
theory exactly the same way, treating Einstein's theory as a gauge theory of 
Lorentz group and making Abelian projection, actually two of them, to obtain 
the restricted gravity. And this restricted gravity presents us a surprising result 
that the graviton could be described (not only by the spin-two metric field but
also) by a photon-like spin-one field.   

To understand this we have to understand the restricted gravity first. To do that 
notice that the above Abelian decomposition is independent of the gauge. 
More importantly the valence part can be viewed as the Lorentz covariant 
gravitational source of the Abelian part. So we can remove the valence part 
without compromising the general invariance, just as we can switch off any 
gravitational source interacting with gravity to obtain the pure Einstein's theory. 
This Abelian projection gives us the restricted gravity. And of course we have 
two restricted gravities, the $A_2$ gravity and $B_2$ gravity.

\subsection{$A_2$ Gravity}

Consider the $A_2$ decomposition first. Clearly (\ref{a2Eeq1}) tells that 
the valence connection $\vZ_\mu$ behaves as the 
Lorentz covariant gravitational source which couple to the restricted connection,
so that we can always put $\vZ_\mu=0$ withut compromising the general invariance 
(or equivalently the Lorentz gauge invariance). Now, with $\vZ_\mu=0$, (\ref{a2Eeq2}) 
is reduced to
\bea
&G_\mn (\pro^\nu \bA^\rho-\pro^\rho \bA^\nu)
-\tG_\mn (\pro^\nu B^\rho-\pro^\rho B^\nu) =0, \nn\\
& \nabla_\mu G^\mn =0,~~~~~\nabla_\mu \tG^\mn =0, \nn\\
& \hat{{\mathscr D}}_\mu \vG^\mn =0.
\label{a2Req1}
\eea
This provides the equations of motion for the $A_2$ gravity. 

To understand the physics behind (\ref{a2Req1}) notice that the first and last 
equations are the first order differential equations, so that they do not 
describe the dynamical (i.e., propagating) graviton. They are the constraint 
equations which determine the connection in terms of the metric. 
But remarkably the two equations for $G_\mn$ and $\tG_\mn$ in the middle 
looks like the free Maxwell's equations. Indeed, since $G_\mn$ and $\tG_\mn$ 
are dual to each other, we can express $G_\mn$ by 
one-form potential $G_\mu$
\bea
&G_\mn=\nabla_\mu G_\nu -\nabla_\nu G_\mu= \pro_\mu G_\nu -\pro_\nu G_\mu,  
\label{a2pot1}
\eea
using the fact $\nabla_\mu \tG^\mn=0$. Equivalently, we can express $\tG_\mn$ by 
one-form potential $\tilde G_\mu$
\bea
&\tG_\mn=\nabla_\mu \tilde G_\nu -\nabla_\nu \tilde G_\mu
= \pro_\mu \tilde G_\nu -\pro_\nu \tilde G_\mu,  
\label{a2pot2}
\eea
using the fact $\nabla_\mu G^\mn=0$. So we can express the equations of 
the restricted metric $G_\mn$ and $\tG_\mn$ (the $\vl$ and $\vtl$ components 
of the Lorentz covariant metric $\vg_\mn$) as a Maxwell-type 
second order differential equation in terms of the potential $G_\mu$,
\bea
& \nabla_\mu G^\mn=0,~~~~~G_\mn= \pro_\mu G_\nu -\pro_\nu G_\mu.
\label{a2Req2}
\eea
This is really remarkable and surprising, because this shows that the dynamical
part of $A_2$ gravity can be described by an Abelian gauge theory. 

\subsection{$B_2$ Gravity}
 
Now, exactly the same way we can have the $B_2$ gravity from the $B_2$ decomposition. 
With $\vZ_\mu=0$, we reduce (\ref{b2Eeq2}) to
\bea
&\cJ_\mn (\pro^\nu K^\rho-\pro^\rho K^\nu)
-\ctJ_\mn (\pro^\nu \tK^\rho-\pro^\rho \tK^\nu) =0, \nn\\
& \nabla_\mu \cJ^\mn =0,~~~~~\nabla_\mu \ctJ^\mn =0, \nn\\
& \hat{{\mathscr D}}_\mu \vG^\mn +\cJ^\mn \hD_\mu \vk-\ctJ^\mn \hD_\mu \vtk=0,
\label{b2Req1}
\eea
which describes the restricted $B_2$ gravity.

Here again the first and last equations can be viewed as the constraint 
equations which determine the connection in terms of the metric. 
But the two equations for $\cJ_\mn$ and $\ctJ_\mn$ in the middle 
allows us to introduce one-form potential $\cJ_\mu$ for $\cJ_\mn$
\bea
&\cJ_\mn= \pro_\mu \cJ_\nu -\pro_\nu \cJ_\mu,  
\label{b2pot1}
\eea
or $\ctJ_\mu$ for $\ctJ_\mn$
\bea
&\ctJ_\mn=  \pro_\mu \ctJ_\nu -\pro_\nu \ctJ_\mu.
\label{b2pot2}
\eea
With this we can express the equations of the restricted metric $\cJ_\mn$ and $\ctJ_\mn$ 
(the $\vj$ and $\vtj$ components of the Lorentz covariant metric $\vg_\mn$) as a Maxwell-type 
second order differential equation in terms of the potential $\cJ_\mu$,
\bea
&\nabla_\mu \cJ^\mn=0,~~~~~\cJ_\mn= \pro_\mu \cJ_\nu -\pro_\nu \cJ_\mu.
\label{b2Req2}
\eea
This shows that the dynamical part of $B_2$ gravity can also be described by an Abelian 
gauge theory. 

Clearly both (\ref{a2Req2}) and (\ref{b2Req2}) imply that the dynamical field
of the restricted gravity is described by a massless spin-one field. But this is 
nothing but the graviton, because the valence part of the Abelian decomposition 
simply becomes a gravitational source of the restricted gravity. This means that 
the graviton can be described by a massless spin-one field. 
This is a most important outcome of our analysis.

At first thought this view sounds heretical, but actually is not so. First of all, 
the massless spin-one field has the right degrees of freedom for the graviton. 
Just as the massless spin-two metric it has two physical degrees. Besides, 
the metric is not the only field which describes the graviton. Classically the metric 
is equivalent to tetrads, so that the graviton can also be described by tetrads. 
And each of the four tetrads becomes a vector. Furthermore, just like the metric, 
our dynamical fields $G_\mn$ and $\cJ_\mn$ are made of tetrads. So it is really 
not a strange idea to describe the graviton by them. The new (and surprising) thing 
of our analysis is that they can be expressed by Abelian potentials, through 
the equation of motion. This leads us to the idea of massless spin-one graviton. 

\section{Discussions}

In this paper we have discussed the Abelian decomposition of
Einstein's theory. Imposing proper magnetic isometries
to the gravitational connection, we have shown how to decompose the
gravitational connection and the curvature tensor into the
restricted part of the maximal Abelian subgroup $H$ of Lorentz group
$G$ and the valence part of $G/H$ component
which plays the role of the Lorentz
covariant gravitational source of the restricted connection, without
compromising the general invariance.

This tells that the Einstein's theory can be viewed as a
theory of the restricted gravity made of the restricted connection
in which the valence connection plays the role of the gravitational
source of the restricted gravity.
We show that there are two
different Abelian decompositions of Einstein's theory,
light-like $A_2$ decomposition (the null decomposition) and 
non light-like $B_2$ decomposition (the space/time decomposition),
because Lorentz group has two maximal Abelian
subgroups.

An important ingredient of the decomposition is the concept of
Lorentz covariant four-index metric tensor $\vg_\mn$ which
replaces the role of the two-index space-time metric $g_\mn$.
We have shown that the metric-compatibility condition
of the connection $\nabla_\alpha g_\mn=0$ is replaced by
the gauge (and generally) covariant condition ${\mathscr D}_\mu \vg^\mn=0$.

From theoretical point of view, the above
decomposition of gravitation differs from the Abelian
decomposition of non-Abelian gauge theory in one important respect.
In gauge theory the fundamental ingredient is the gauge potential,
and the decomposition of the potential provides
a complete decomposition of the theory.
But in gravitation the fundamental field is
assumed to be the metric, not the connection (the potential).
Because of this the decomposition of the connection gives us the
the decomposition of the metric only indirectly,
through the equation of motion. 
It would be very interesting to see if one can actually
decompose the metric explicitly, and decompose the Einstein's theory
in terms of the metric.

Nevertheless the above decomposition of Einstein's theory has deep
implications. First of all, this tells that we can construct
a restricted theory of gravitation, actually two of them, which is generally 
invariant (or equivalently Lorentz gauge invariant) but has fewer physical 
degrees of freedom than what we have in Einstein's theory. 
This means that we can separate the Abelian part of gravity which describes 
the core dynamics of Einstein's theory without compromising
the general invariance. More importantly, our analysis shows that we could 
describe the restricted gravity by an Abelian gauge theory 
with one-form potential. In other words, our result implies 
that the graviton can be described by a massless spin-one potential, in stead of 
the spin-two metric. This has a very important implication, because this point 
can play a crucial role for us to construct the quantum gravity. 

Furthermore, the decomposition makes the topology of Einstein's
theory more transparent. Indeed with the Abelian decomposition we
can study the topological structures of the theory more easily,
because the topological characteristics are imprinted in the
magnetic symmetry. For example, the $A_2$ decomposition makes it
clear that the topology of Einstein's theory is closely related to
the topology of $SU(2)$ gauge theory. This is natural, because
$SU(2)$ forms a subgroup of Lorentz group. This similarity between 
Einstein's theory and $SU(2)$ gauge theory might be very useful for us 
to study the gravito-magnetic monopole in Einstein's theory which has 
the monopole topology $\pi_2(S^2)$ \cite{cho91,new}.

Perhaps more importantly, this strongly implies that Einstein's
theory may have the multiple vacua similar to what we find in
$SU(2)$ gauge theory. This turns out to be true. In fact with a
proper magnetic isometry we can construct all possible vacuum
space-times, and show that Einstein's theory has exactly the same
multiple vacua that we have in $SU(2)$ gauge theory which can be
classified by the knot topology $\pi_3(S^3)=\pi_3(S^2)$
\cite{grg0}. 

This could have a far reaching consequence. Just as in
$SU(2)$ gauge theory, the multiple vacua in Einstein's theory can
be unstable against quantum fluctuation. And there is a real
possibility that Einstein's theory may admit the
gravito-instantons which can connect topologically distinct vacua
and thus allow the vacuum tunnelling \cite{grg0,egu}. Clearly this
will have an important implication in quantum gravity.

The details of the subject with
interesting applications will be discussed separately \cite{grg3}. 

{\bf ACKNOWLEDGEMENT}

~~~The work is supported in part by Korea National Research Foundation 
(Grants 2007-314-1-C00055, 2008-314-1-C00069, and 2010-002-1564) and by Ulsan National 
Institute of Science and Technology.

\end{document}